\documentclass[preprint,12pt]{elsarticle}

\usepackage{amssymb}
\usepackage{tabularx}
\usepackage{pifont} % For \ding
\newcommand{\xmark}{\ding{55}}
\newcommand{\cmark}{\ding{51}}
\graphicspath{{./figures/}}
\DeclareGraphicsExtensions{.pdf}
\usepackage{amsmath,xcolor}
\usepackage{algorithmicx}
\usepackage{algpseudocode}
\usepackage{subfig}
\usepackage{graphicx}
\usepackage{comment}
\usepackage{url}
\usepackage[ruled]{algorithm}
\definecolor{light-gray}{gray}{0.75}
\algrenewcommand{\algorithmiccomment}[1]{\hskip3em{{\footnotesize \textcolor{light-gray}{$\blacktriangleright$}}} #1}
\usepackage{xspace}

\usepackage{hyperref}

\newcommand{\tdblite}{AerialDB\xspace}
\newcommand{\comm}[1]{}

\newcommand{\ysnote}[1]{ {\textcolor{magenta} { ***Yogesh: #1 }}} % needs a response
\newcommand{\ssnote}[1]{ {\textcolor{blue} { ***Subhajit: #1 }}} % needs a response

\newcommand{\ysnoted}[1]{ {\textcolor{green} { ***TODO Later: #1 }}} % postpone addressing of comment
\renewcommand{\ysnoted}[1]{}

\usepackage[normalem]{ulem} % required for strikeout font

\newcommand{\modc}[1]{{\textcolor{blue}{#1}}}
\newcommand{\addc}[1]{{\textcolor{teal}{#1}}}
\renewcommand{\addc}[1]{#1}
\renewcommand{\modc}[1]{#1}
\newcommand{\modcb}[1]{{\textcolor{blue}{#1}}}
\newcommand{\addcb}[1]{{\textcolor{teal}{#1}}}
\renewcommand{\addcb}[1]{#1}
\renewcommand{\modcb}[1]{#1}

\newcommand{\delc}[1]{ {\textcolor{gray} {\sout{#1}} }}
\renewcommand{\delc}[1]{}

\usepackage{blindtext}

\journal{Pervasive and Mobile Computing}

\begin{document}

\begin{frontmatter}



\title{AerialDB: A Federated Peer-to-Peer Spatio-temporal Edge Datastore for Drone Fleets}

% \affiliation[label1]{organization={},

% \affiliation[label2]{organization={},

\author[uiuc]{Shashwat Jaiswal}
\author[iisc]{Suman Raj}
\author[iitkgp]{Subhajit Sidhanta}
\author[iisc]{Yogesh Simmhan}%% Author name

\affiliation[uiuc]{organization={University of Illinois Urbana-Champaign},%Department and Organization
            % city={},
            state={IL},
            country={USA}}
\affiliation[iisc]{organization={Indian Institute of Science},%Department and Organization
            % city={},
            % state={IL},
            country={India}}
\affiliation[iitkgp]{organization={Indian Institute of Technology Kharagpur},%Department and Organization
            % city={},
            % state={IL},7
            country={India}}
%% Abstract
\begin{abstract}
% \ysnote{BeeStore, TejDB, MeaDB, DataComb, Apiary, ApisDB, HiveStore, AerialDB, FlyStore}
% \srnote{AerialDB or FlyStore?} \shashwat{I like BeeStore}
% \ysnote{@SS to take a pass}
 Recent years have seen an unprecedented  explosion in research that leverages the newest computing paradigm of Internet of Drones comprised of a fleet of connected Unmanned Aerial Vehicles (UAVs) used for a wide range of tasks such as monitoring and analytics in highly mobile and changing environments characteristic of disaster regions. % has received a tremendous impetus
% Due to their innate mobility and versatility, state-of-the-art UAVs possess the  unique ability to provide the above services to regions where such services would not be  practically possible to deliver by traditional means. 
% Further, the meteoric rise in cutting-edge research in the area of cyber-physical systems has made it possible for UAVs to serve the purpose of acting as a computing infrastructure to process the data collected from on-board cameras as well as ground-based IoT (Internet of Things) devices, such as various kinds of sensors. 
 Given that the typical data (i.e., videos and images) collected by the fleet of UAVs deployed  in such scenarios can be considerably larger than what the onboard computers can process, the UAVs need to offload their data in real-time to  the edge and the cloud for further processing. 
% manner, the input data or its associated metadata needs to be stored and processed locally, i.e., on the UAVs themselves.  
To that end, we present the design of \tdblite - a lightweight decentralized data storage and query system that can store and process time series data on a multi-UAV system comprising: A) a fleet of \modc{hundreds} of UAVs fitted with onboard computers, and B) ground-based edge servers connected through a cellular link. Leveraging lightweight techniques for content-based replica placement and indexing of shards, \tdblite has been optimized for efficient processing of different possible combinations of typical spatial and temporal queries performed by real-world disaster management applications. Using containerized deployment spanning up to 400 drones and 80 edges, we demonstrate that \tdblite is able to scale efficiently while providing near real-time performance with different realistic workloads. Further, \tdblite comprises a decentralized and locality-aware distributed execution engine which provides graceful degradation of performance upon edge failures with relatively low latency while processing large spatio-temporal data. \modc{\tdblite exhibits comparable insertion performance and 100 times improvement in query performance against state-of-the-art baseline. Moreover, it experiences a 10 times improvement in performance with insertion workloads and 100 times improvement with query workloads over the cloud baseline.}
\end{abstract}


% \begin{highlights}
% \item A reliable federated spatio-temporal datastore that uses lightweight content-based replica placement and indexing of shards, onto InfluxDB services running on the edge, for data sourced from mobile devices (drones) onto wide-area distributed edge servers.
% \item A decentralized and locality-aware distributed execution engine for processing spatio-temporal queries from query clients with load-balancing and graceful degradation of performance upon edge failure.
% \item Detailed experiments to evaluate the insertion, query, and reliability performance of \tdblite by emulating up to 400 drones and 80 edge devices using containers, and comparing it against a cloud-only baseline.
% \item 10 times improvement in performance with insertion workloads and 100 times improvement with query workloads over the cloud baseline is demonstrated.
% \end{highlights}

%% Keywords
\begin{keyword}
Distributed Systems \sep Spatio-Temporal Databases \sep Drones \sep Edge Computing \sep Mobile Computing \sep Multi-UAV Systems
%\srnote{Multi-UAV Systems and Drone Fleets seem redundant, we can skip one of them?}



\end{keyword}

\end{frontmatter}



\section{Introduction}
\label{intro}
% \Note{1.5pgs}
% \ysnote{SS to take a pass, followed by YS}
% \ssnote{Finished pass. YS  may please take over}

%%-----------------------------------------------------
%%% Drones as sources of spatio-temporal data streams
\paragraph{Fleets of UAVs as Data Collection Platforms}
We are faced with an unprecedented increase in the frequency of occurrence of natural disasters over the past few years which are affecting our health, economy, infrastructure, and causing fatalities.  This necessitates ready availability of near real-time computing platforms for hosting various emergency forewarning, disaster prevention, and survivor tracking applications for performing  monitoring, prediction, and actuation tasks that are essential for managing such disaster situations. As such, the paradigms of pervasive computing and mobile computing are showing tremendous promise in addressing many challenges faced by traditional computing platforms in such scenarios.  Specifically,  Unmanned Aerial vehicles (UAVs), also called drones, have become the most commonly used platforms  because of  innate maneuverability, versatility, and  ability to support computation, communication, and sensing on demand \cite{9665659}. They are used for performing various computational tasks like real-time data collection, disaster monitoring, providing logistics support in disaster scenarios~\cite{7807176}, urban safety~\cite{noor2018remote}, traffic and infrastructure monitoring~\cite{9128519,ham2016visual}, mapping agricultural and forest land~\cite{farmbeats}, etc. 
\par Fleets of UAVs can also act as first-class sensing platforms, collecting large troves of observational data across city-scale regions, rendering them optimally suited for tasks related to disaster management such as damage estimation and recovery planning. Such fleets of UAVs, also called Internet of Drones, are used as the platform for deploying such smart city (or smart village)  applications, carrying diverse sensors, such as visual, thermal and hyper-spectral cameras~\cite{ASADZADEH2022109633}, air quality and gas detectors~\cite{ROHI2020e03252,s22030860}, weather and noise monitors~\cite{SZIROCZAK2022100859,9773134}, electronics and communication sniffers~\cite{8598647}, etc. These sensors generate a stream of time-series observations while the UAVs perform various critical emergency missions around the city (or village) ~\cite{10099247, 10.1145/3307334.3328627, DBLP:conf/infocom/KhochareSS021}. Such observations can even be collected as an ancillary part of some other activity, such as ferrying packages ~\cite{moninger2003automated}. This gives rise to a large trove of real-time spatio-temporal data that can be put to use for novel applications such as traffic management, air, and noise pollution estimation, cellular signal-strength estimation, etc. Such requirements even extend to crowd-sourced data collection through smartphones and other mobile devices. \addc{Effectively managing and querying this data requires a spatiotemporal database. A spatiotemporal database is defined as a database designed to store, manage, and query spatial data that evolves over time, integrating both location and temporal dimensions~\cite{pant2018survey,abraham1999survey}.}
% restas2015drone

%% need for distributed database: on-board analytics, offload in realtime to cloud, offload when they reach fleet depot. Consumers may be in the city as well, so cloud may not be beneficial. Backhaul network to cloud may also be low bandwidth/costly. So makes sense to retain the data within the MAN. 5G is also bringing Telco Edges on-board. Can we use these edge devices for maintaining distributed data and query upon them? Others have done distributed block stores and files, but not spatio-temporal querying.

%%-----------------------------------------------------
%%% telco edge/fog to support drones
\paragraph{Limitation of the Cloud as Data Storage Platforms}
The traditional data management paradigm is concerned with pushing the above data to the Cloud with the goal of safely storing it (data) 
 such that it can be subsequently retrieved in future in near real-time. This can be done using a backhaul network  from the fleet depot of the drones (UAVs) when the drones land there, or using wireless/cellular networks when they are flying. However, in disaster regions where the network connectivity is severely compromised, it would be practically infeasible to acquire a stable internet connection to be able to store/retrieve data to/from the cloud. 
\par Further, applications may require near real-time access to the spatio-temporal observations, motivating the need for offloading data from the drones as and when they are collected. Also, the clients that would query the observations are likely to be in local proximity, near the drones, or at least within the Metropolitan Area Network (MAN) of the city (or village) while being far away from the cloud. With traditional cloud databases, this adds to the data access latency/variability and the backhaul network cost -- from the drones to the clients through the cloud, which may not be acceptable for applications like urban safety, traffic management, and where the drone's navigation depends on the outcome of the data analysis.  Lastly, emergency and disaster situations may curtail network connectivity,  or disconnect access to the cloud completely. These factors push the need to revisit a cloud-centric approach to spatio-temporal data management for drones.

%%-----------------------------------------------------
%%% telco edge/fog to support drones
\paragraph{Opportunity for Data Management on the Edge:} Drone fleets are supported by base stations on the ground for communications and computation, e.g., using 4G and 5G cellular links for the network link, telco edge from 5G providers, and edge platforms from Content Distribution Networks (CDN) providers like Akamai~\cite{akamai}. Such edge \modc{servers}, also called \textit{fogs}, are present within the MAN, in spatial and network proximity of the drones and their data consumers. Their communication link allows navigation controls to be sent to the drones and data to be offloaded from them, while their computing platform allows hosting of services and storage.

Such distributed edge \modc{servers} with the capacity of low-end workstations or servers offer a natural platform for management and querying over such observational data across the MAN. Being usually available in large numbers and spread across the city, they offer access to the data within one or a few hops with low latency for drones to offload and clients to query. This also facilitates replication of the data for resiliency, preventing a single point of failure, and scalable query execution across the distributed edge \modc{servers}. Downstream applications that consume the output of the queries can also be co-located on the same edge \modc{servers} to avoid further data movement.

\paragraph{Gaps in the Literature:}
% \ysnote{lot of work on task offloading, not data offloading and management for mobile devices}
% \ysnote{distributed block storage on edge resources, but not higher order querying of spatial and temporal}
%\ysnote{@Suman to revisit}
The offloading of tasks from mobile platforms to edge and cloud resources for processing has been well-studied~\cite{neumann2011stacee, 10.1145/3447993.3448628}. However, the management of data offloaded from such devices on distributed edge \modc{servers} is less explored. There is some work on storing and managing files on federated edges~\cite{yuan2021csedge, 8956055, xia2019secure, sonbol2020edgekv, qiao2020trustworthy, nicolaescu2021store, monga2019elfstore, linaje2019mist}, drone-based applications require support for spatio-temporal queries. 
%\ysnote{downsides of torquedb?}
 The closest baseline among the prior systems is TorqueDB \cite{garg2020torquedb}, which processes time series queries on the edge, does not leverage the spatial and temporal dimensions for any placement decisions, thus preventing it from leveraging spatial and temporal locality, and adapting to the mobility of the drone.

% \ysnote{Limited support for distributed timeseries databases, even in the cloud.}
Contemporary time-series and spatio-temporal databases such as IoTDB and InfluxDB themselves are limited to single-machine deployments rather than distributed execution on the cloud. This limits their scalability, and
% \ysnote{data systems with centralized elements cause bottlenecks and single points of failure. A fully federated P2P approach with decentralized discovery.}
also renders these databases highly susceptible to single point of failure in case the cloud VM (i.e., Virtual Machine) or the network link to it is lost. 
% Such a loss is exacerbated if single-instance data stores on edge resources also makes them prone to failures and loss of data availability.

  %%% Some related work on edge and fog storage and our benefits over them
%\ssnote{@Shashwat and @Suman kindly verify the following claim against the cited papers.}
Existing literature on edge and fog data management are not specialized for spatio-temporal data~\cite{tao2002time,monga2019elfstore}, mobility of the edge devices~\cite{kolomvatsos2019spatio, chen2018spatio, garg2020torquedb}, fully decentralized operations~\cite{aral2018decentralized}, and/or the use of edge and fog as the primary data plane~\cite{linaje2019mist}. 
 % \ysnote{Suman/Shashwat, can you take a stab at improving this para with relevant/recent literature on edge/fog storage and drone-based data management to contrast against ours?}

%%-----------------------------------------------------
%%% What are we proposing
\paragraph{Contributions:}
We propose \textit{\tdblite} - a distributed data platform to store and query over spatio-temporal data collected from mobile platforms such as drones in a decentralized, performant, scalable, and reliable manner using edge \modc{servers}. The system architecture avoids any single point of failure, rendering them perfectly suited for providing near real-time response in emergency use cases such as natural disasters. We design content-based addressing and replica placement strategies over the spatial and temporal properties of the data to enable federated data discovery, data locality, and resilience. We also develop an indexing mechanism over spatio-temporal features to scope the search space, as well as distributed queries across relevant edges holding the data. Within each edge, \tdblite uses InfluxDB to perform the actual query, helping it support a rich set of query predicates offered by this contemporary (single-machine) database. \addc{To the best of our knowledge, \tdblite is the first federated spatio-temporal datastore for mobile nodes.}
% a  it can cover most typical geospatial temporal queries including the  case of traditional cloud computing. It is also designed to adapt to diverse observational workloads from mobile edge sources.

We make the following specific contributions in this paper.
\begin{itemize}
      \item We design \tdblite, a reliable federated spatio-temporal datastore (\S~\ref{sec:design}) that uses lightweight content-based replica placement and indexing of shards (\S~\ref{sec:arch:insert}, \S~\ref{sec:arch:insert:hash}) onto InfluxDB services running on the edge, for data sourced from mobile devices (drones) onto wide-area distributed edge servers.
      \item We develop a decentralized and locality-aware distributed execution engine for processing spatio-temporal queries from query clients (\S~\ref{sec:arch:query}) with load-balancing and graceful degradation of performance upon edge failure (\S~\ref{sec:arch:query:resilience}).
      % \item A lightweight content based indexing and replica placement scheme
      % \item A fast temporal and spatially local query resolution mechanism. (in the presence of mobility)
      % \item A mechanism to ensure reliability against device failures.      
      \item We implement this design and perform detailed experiments to evaluate the insertion, query, and reliability performance of \tdblite by emulating up to 400 drones and 80 edge devices using containers, and compare it against a cloud-only baseline (\S~\ref{sec:results}). We demonstrate a 10 times improvement in performance with insertion workloads and 100 times improvement with query workloads over the cloud baseline. %\Note{Our results indicate ???\% better performance} .
      % We demonstrate that \tdblite  beats the cloud baseline by x\%.
\end{itemize}

%\ysnote{Emphasize novelty on content based addressing on different dimensions, over-replicated indexing for decentralized search....}

\section{Related Work}
% \Note{1pgs}
% \ysnote{SS to take a pass and hand over to YS}
% \ssnote{Seems fine to me. some very minor updates done. YS may take over please.}
% \ysnote{Categories:\\
% Edge/fog storage (like torqueDB, Eyal/UToronto)\\
% Spatio-temporal DB: centralized and distributed\\
% Data store and mobility\\
% Offloading of tasks and data\\
% }

% \srnote{mention elfstore and torquedb in 3rd person}

\subsection{Edge/Fog Storage and Querying}
With the advent of the edge computing paradigm, it is crucial to host data storage and querying services at the edge \addc{for low-latency, resilient, and real-time spatiotemporal query processing, especially in connectivity-constrained environments like disaster zones}~\cite{nicolaescu2021store}~\cite{yuan2021csedge}~\cite{sonbol2020edgekv}. It becomes challenging in deployments where multiple edge and fog devices offer services for scalability and fault tolerance. ElfStore~\cite{8818408} proposes a resilient data storage service for stream-based, block-oriented distributed storage service over unreliable edge devices, \addc{but lacks native support for querying, making it unsuitable for applications requiring real-time spatiotemporal data analysis.}. 

Linaje et al. \cite{linaje2019mist} present a novel data
distribution and replication storage solution for wireless sensor networks using cheap storage devices hosted on the sensor nodes, \addc{but their approach is limited to static nodes and does not address mobility or distributed spatiotemporal querying.}
TorqueDB~\cite{10.1007/978-3-030-57675-2_18} proposes a distributed query engine over time-series data that operates on edge and fog resources, \addc{but it does not consider spatial properties for query optimization, thereby missing spatiotemporal locality benefits crucial for mobile applications}. Xia et al. \cite{xia2019secure} present RoSES - a robust security-aware data storage model based on TLRC - a new variant of Erasure codes that enables lightweight computation on secure data at the edge. Qiao et al.~\cite{qiao2020trustworthy} \modc{use reinforcement learning for edge storage optimization, yet their work is specific to intelligent transport systems and lacks a generalized spatiotemporal query execution model}. \delc{In contrast with the above, we propose a framework that supports mobile data storage as well as data querying on edges.} \addc{In contrast, our framework integrates mobile data storage with efficient spatiotemporal query execution, ensuring real-time processing, fault tolerance, and adaptability to mobility, making it uniquely suited for dynamic environments like disaster response and UAV-based analytics.} 
% Further, RoSES is integrated with TODA - an adaptable data access strategy that allows the processing of  legitimate, secure data requests from uncertain users.  

% there is a pressing need to store and query data at the edge.   

\subsection{Data Store and Mobility}
As the Internet-of-Things (IoT) ecosystem is rapidly gaining popularity, drones have emerged as mobile sensing devices for outdoor environments~\cite{10.1145/3307334.3328627}~\cite{10.1145/3447993.3483273}. Big Data on the Fly~\cite{9166731} propose UAV-mounted data sensing for disaster management.  This, however, opens up multiple challenges in determining the location of drones to maximize data collection utility~\cite{9439126}. Towards this, BeeCluster~\cite{10.1145/3386901.3388912} enables developers to express a sequence of geographical sensing tasks and map them to a fleet of drones using predictive optimization. 

STACEE~\cite{neumann2011stacee} discusses the creation of a storage cloud using edge devices based on Peer-to-Peer resource provisioning while minimizing energy. MobilityDB~\cite{10.1145/3406534} is a moving object database that offers an SQL query interface with abstract data types for representing moving object data. Motivated by this, we incorporate the ability to handle mobility in three primary components of our data store: collection, insertion, and querying.   

% The variety of data collected by heterogeneous 

\subsection{Offloading of tasks and data}
Data collected by drones may be required to offload to nearby powerful edges for compute-intensive tasks, which can be heterogeneous in nature~\cite{10171496}. Elf~\cite{10.1145/3447993.3448628} proposes a framework to accelerate mobile deep vision applications by partitioning the video frame and offloading partial inference tasks to multiple servers for parallel processing. Oikawa et al. propose an efficient density-based data selection and management method for edge computing~\cite{9439127}. This motivates the need to intelligently place data and tasks in edge computing environments~\cite{8767386}. 

Busacca et al.~\cite{busacca2020drone} propose a game theoretical approach for providing optimized resources to users in a UAV-assisted edge computing scenario. In contrast to our proposed system, this work presents a theoretical cost model that is not translated into actual implementation. Yu et al.~\cite{8956055} propose a joint task offloading and resource allocation problem for UAV-enabled systems to minimize latency and energy consumption. 
% where users leverage drones as edge devices for network connectivity, computation, and storage in resource-constrained situations. 
 Jeong et al.~\cite{jeong2017mobile} present an approach for energy optimization using joint optimization of UAV trajectory and the bit allocation in uplink and download communication between UAVs and mobile users in UAV-hosted cloudlets. We, however, leverage the concept of parent edge, which lies in the spatial partition of the current location of the drone and triggers the data storage pipeline.  

% , such as running a Deep Neural Network (DNN) for video analytics. However, for smart city applications, edge servers may be heterogeneous in nature and may require an 

\subsection{Spatio-temporal DB: centralized and distributed}
Since we focus on data collection by mobile drones, we need a lightweight framework that can store and query spatio-temporal data. Pelekis et al.~\cite{pelekis_theodoulidis_kopanakis_theodoridis_2004} provide an extensive review of the early spatiotemporal database models. Koubarakis et al.~\cite{koubarakis2012teleios} propose an implementation of scientific database and semantic web technologies to store satellite images linked with geospatial data.  ChronosDB~\cite{zalipynis2018chronosdb} is a geospatial scalable database that offers a wide variety of array operations at scale by partially delegating in-situ processing to tools that are optimized for a single machine. Apache IoTDB~\cite{10.1145/3589775} is a time-series database management system for IoT applications. 

Feather~\cite{9355826} offers a geo-distributed, hierarchical, eventually consistent tabular data store that supports efficient global queries using a flexible temporal freshness guarantee by utilizing the hierarchical structure of edge networks. In contrast with the existing literature, our goal is the development of a lightweight storage system that performs most of the query processing tasks on edges efficiently, without relying on a centralized cloud or a requirement for a stable network. 

\subsection{Dynamic partition of data}
It is well established that a dynamic partitioning strategy is essential to enable a system, especially for indexing a storage system, to adapt to the changing nature of the data processed and the workload executed. Dynamic partition of metadata across servers has been explored in past research but has not yet been solved entirely.  Weil et al.~\cite{weil2004dynamic} have presented an indexing technique that stores the intermediate metadata in the form of a bounded log structure, and thus partitions a hierarchical dataset into several subtrees. Hermes~\cite{nicoara2015hermes} is a system in which the representative graph is of smaller size, such that a partitioning algorithm like Kernighan-Lin can be applied. Zhao et al.~\cite{zhao2016toward} propose an indexing scheme that leverages Zero-hop distributed hash table (ZHT) to provide a scalable dynamic indexing service that is able to handle churns. Contrastingly, the system design presented in this paper is based on a simple yet effective lightweight dynamic partitioning approach.

\begin{table}[t]
\centering
\caption{\addcb{Comparison of Related Works Based on Key System Properties}}
\label{tab:related_works}
% \resizebox{\textwidth}{!}{%
\begin{tabular}{|l|c|c|c|c|}
\hline
\textbf{System} & \shortstack{\textbf{Spatio-}\\\textbf{Temporal}\\\textbf{Queries}} & \shortstack{\textbf{Mobility}\\\textbf{Aware}} & \shortstack{\textbf{Decen-}\\\textbf{ralised}} & \shortstack{\textbf{Fault}\\\textbf{Tolerant}}\\
\hline
TorqueDB \cite{garg2020torquedb}     & \xmark  & \xmark     & \cmark & \cmark  \\
ElfStore \cite{monga2019elfstore}    & \xmark  & \xmark     & \cmark & \cmark  \\
Feather \cite{9355826}               & \xmark  & \xmark     & \cmark & \xmark  \\
MobilityDB \cite{10.1145/3406534}    & \cmark  & \cmark     & \xmark & \xmark  \\
InfluxDB    & \cmark  & \xmark     & \xmark & \xmark  \\ \hline
\textbf{\tdblite (Ours)}              & \cmark  & \cmark     & \cmark & \cmark  \\
\hline
\end{tabular}
% }
\end{table}

\subsection{\addcb{Summary}}
\addcb{To better contextualize our contributions, we reflect on the limitations of existing systems in relation to the unique demands of spatio-temporal data management for mobile drone fleets. Table~\ref{tab:related_works} summarizes the key dimensions that distinguish \tdblite's integrated design from the more siloed capabilities of existing systems.
While several works address edge-based storage \cite{monga2019elfstore, 9355826}, distributed time-series querying \cite{garg2020torquedb}, or mobility-aware design \cite{10.1145/3406534}, few integrate spatio-temporal querying with support for indexing to speed up queries, mobility-tolerant edge querying that adapts to changing network connectivity, decentralized spati-temporal replica placement and distributed querying, and fault resilience in a unified system.
\tdblite is unique in leveraging lightweight content-based replica placement and indexing using spatial, temporal, and identifier-based hash functions, while also ensuring graceful performance degradation under edge failures.}

%\newpage
\comm{\subsection{Earlier Text on Related Work}
Yu et al. \cite{yu2021geosparkviz} propose a system to scale geospatial queries and co-optimize the query execution and visualization of spatial maps on traditional servers. 
% \srnote{not sure if relevant, this paper is more on co-optimising query execution and visualization of spatial data, and claims to outperform existing literature for visual analytics tasks.}
Zalipynis et al.~\cite{zalipynis2020bitfun} proposes a novel approach of bitmap-based re-indexing of geospatial queries while executing queries with similar tunable functions to achieve a significant improvement in query performance.
% \srnote{we perform indexing during insertion, but they are doing it during querying and this is for optimizing queries with similar mathematical functions, relevant??}  
Koubarakis et al. \cite{koubarakis2012teleios} proposes an implementation of scientific database and semantic web technologies on top of MonetDB to store satellite images linked with geospatial data. 
% \srnote{can be used in the introduction, as it addresses the need for scalable access to petabytes of geospatial data (virtual Earth Observatory)}
Zalipynis et al. \cite{zalipynis2018chronosdb} presents ChronosDb a scalable database that performs in-situ processing of geospatial data files to produce improved scaling and performance improvement. 
% \srnote{ChronosDB offers a wide variety of array operations at scale by partially delegating in-situ processing to tools that are optimized for a single machine.}
Kazemitabar et al. \cite{kazemitabar2010geospatial} proposes leveraging Microsoft StreamInsight to perform optimized processing of geospatial stream queries. 
% \srnote{shows a demonstration on transportation data}
Rodriges et al. \cite{rodriges2022webarraydb} proposes WideArrayDb - a database which can return the result of geospatial queries on the web. 
% \srnote{the first geospatial array DBMS that runs completely inside a web browser, relates to edge computing?} 

 Apache Spark and Spark-based systems have various significant features that enable them to store very large spatial data  \cite{yu2019geospatial}. 
 % \srnote{Support of Spark for spatial data, can shift above to introduction probably?}
 Kefaloukos et al. \cite{kefaloukos2014declarative} propose a declarative language to perform cartographic operations within relational databases that store geospatial data. 
 % \srnote{More on visualization, not relevant??}
 Patel et al. \cite{patel1997building} is one of the earlier works that demonstrates the effect of parallelizing spatial database operations on object relational database systems. 
 % \srnote{scalable geo-sptial DBMS}
 Dong et al. \cite{dong2020marviq} presents an approach for efficient spatial visualization of  geospatial range queries based on a scheme of partitioning the selection-attribute domain into smaller intervals. 
 Rueda et al.  \cite{rueda2006extensible} proposes a web service based approach of composing different tools and resources for processing geospatial data collected from diverse repositories. Similarly Jaegar et al.  \cite{jaeger2005scientific} proposes an extensible framework to process data streams using remote sensing pipelines spanning multiple nodes in a network.  Prasher et al. \cite{prasher2003efficient} proposes a novel approach of retrieval of data stored at different resolution levels in  geospatial databases. Hariharan et al. \cite{hariharan2007processing} proposes a novel framework for efficient processing of spatial key queries on GIR databases. Shaw et al. \cite{shaw2007efficient} presents design of a novel indexing scheme for efficient execution of spatial network queries. Wiegand et al. \cite{wiegand2003web} presents a web based system for efficient processing of geospatial database queries. To the best of our knowledge \tdblite is the first of its kind database system for efficient execution of geospatial timeseries queries in the edge.

%% relevant citations from mobicom
In-Situ Data Curation: A Key to Actionable AI at the Edge~\cite{10.1145/3495243.3558758}\\
% Visage: Enabling Timely Analytics for Drone Imagery~\cite{10.1145/3447993.3483273}\\
% Elf: accelerate high-resolution mobile deep vision with content-aware parallel offloading~\cite{10.1145/3447993.3448628}\\
Hermes: an efficient federated learning framework for heterogeneous mobile clients~\cite{10.1145/3447993.3483278}\\
Flexible high-resolution object detection on edge devices with tunable latency~\cite{10.1145/3447993.3483274}\\

%% relevant citations from mobisys
% Data Collection from Outdoor IoT 802.15.4 Sensor Networks using Unmanned Aerial Systems (poster)~\cite{10.1145/3307334.3328627}\\
TEO: ephemeral ownership for IoT devices to provide granular data control~\cite{10.1145/3498361.3539774}\\
FedBalancer: data and pace control for efficient federated learning on heterogeneous clients~\cite{10.1145/3498361.3538917}\\
Low-latency speculative inference on distributed multi-modal data streams~\cite{10.1145/3458864.3467884}\\
% BeeCluster: drone orchestration via predictive optimization~\cite{10.1145/3386901.3388912}\\
IDEA: intelligent divine eye on air through multi-UAV collaborative inference~\cite{10.1145/3498361.3538674}\\
FlyZone: A Testbed for Experimenting with Aerial Drone Applications~\cite{10.1145/3307334.3326106}\\

%% relevant citations from percom
DroneVLC: Exploiting Drones and VLC to Gather Data from Batteryless Sensors~\cite{10099247}\\
Zone-based Federated Learning for Mobile Sensing Data~\cite{10099308}\\
Reservoir: Named Data for Pervasive Computation Reuse at the Network Edge~\cite{9762397}\\
SmartSPEC: Customizable Smart Space Datasets via Event-driven Simulations~\cite{9762405}\\
Consent-driven data use in crowdsensing platforms: When data reuse meets privacy-preservation~\cite{9439125}\\
% Data Collection Utility Maximization in Wireless Sensor Networks via Efficient Determination of UAV Hovering Locations~\cite{9439126}\\
% Density-Based Data Selection and Management for Edge Computing~\cite{9439127}\\
ERAIA - Enabling Intelligence Data Pipelines for IoT-based Application Systems~\cite{9127385}\\
% Context-Aware Data and Task Placement in Edge Computing Environments~\cite{8767386}\\
bioSmartSense: A Bio-inspired Data Collection Framework for Energy-efficient, QoI-aware Smart City Applications~\cite{8767392}\\

 \par There have been a few state-of-the-art systems research prototypes that  aim at providing data storage service at the edge \cite{nicolaescu2021store} \cite{yuan2021csedge} \cite{sonbol2020edgekv}. However, to the best of our knowledge, there is a general lack of research aimed toward addressing the research challenges specific to developing storage on drones \cite{yang2019energy}. 
 Busacca et al.  \cite{busacca2020drone} propose a game theoretical approach for providing optimized resources to users in a UAV-assisted edge computing  scenario where users leverage drones as edge devices for network connectivity, computation,  and storage in resource-constrained situations. As opposed to our proposed system, this work presents a theoretical cost  model that is not translated into  actual implementation.
 Garg et al. \cite{garg2018uav} propose a triple bloom filter-based approach for fast processing of data/service requests from vehicles in an ITS (intelligent transport system)  network in UAVs for enabling real-time processing  of critical data in a secure manner. 
 Jeong et al. \cite{jeong2017mobile} present an approach for energy optimization by the way of joint optimization of UAV trajectory and the bit allocation in  uplink and download communication between UAVs and mobile users in UAV-hosted cloudlets.
 Linaje et al. \cite{linaje2019mist} present an approach to store the data from sensor networks using cheap storage devices hosted on the sensor nodes. In contrast with the cloud and the fog, the above model of storage in the edge/mist incurs much 
 % less  cost while not compromising the privacy of data.
 Neumann et al. \cite{neumann2011stacee} discuss an architecture where the mobile edge 
 devices can be used to contribute towards the storage capability of storage clouds such that energy consumption and latency are minimized. 
 Nicolaescu et al. \cite{nicolaescu2021store} present the design of SEND - a prototype storage framework that provides persistence of data at the network edge comprised of various edge devices. Based on a system-wide identifier of the data called labels, SEND strategically places the data taking into account factors such as the data source, proximity to the data processing functions, etc. 
 Nyamtiga et al. \cite{nyamtiga2019blockchain}  and Ren et al.  \cite{ren2019secure} present an architectural solution to incorporate edge computing for supporting the deployment of blockchain-based decentralized P2P solutions for IoT applications  such as smart  healthcare leveraging storage capabilities of the edge for transactional support.
 Sondur  et al. \cite{sondur2019storage} evaluate commercially available edge storage infrastructure implementations and suggest some design-level tuning that must be performed on the same to support the real-time processing of IoT workloads. 
 Xia et al. \cite{xia2019secure} presents RoSES - a robust security-aware data storage model based on TLRC - a new variant of Erasure codes that enables lightweight computation on secure data at the edge. Further, RoSES is integrated with TODA - an adaptable data access strategy that allows the processing of  legitimate, secure data requests from uncertain users. 
 Yang et al. \cite{yang2020multi} present a deep reinforcement learning-based approach to load-balancing the computation on a multi-UAV system that  acts as a mobile edge computing platform. 
 Zhou et al.  \cite{zhou2018computation} consider the research problem of joint optimization of computational efficiency and energy in a UAV-hosted MEC setup. 
 Qiao et al. \cite{qiao2020trustworthy} propose a reinforcement learning-based technique for optimizing the storage available in edge nodes in an intelligent transport system.  
  \ssnote{@Shashwat and Suman please see if we can add some more analysis here.}
In contrast with the above papers, our goal is the development of a lightweight storage system that performs most of the query processing tasks on drones itself efficiently, without relying on a centralized cloud or any requirement for a stable network. 
\par It is well established that a dynamic partitioning strategy is essential to enable a system, especially for indexing a storage system, to adapt to the changing nature of the data processed and the workload executed. Dynamic partition of metadata across servers has been explored in past research but has not been yet solved.  Weil et al. \cite{weil2004dynamic} have presented an indexing technique that stores the intermediate metadata in the form of a bounded log structure, and thus partitions a hierarchical dataset into several subtrees.  Nicoara et al. \cite{nicoara2015hermes} present Hermes - system in which the representative graph is of smaller size, such that a partitioning algorithm like that of Kernighan-Lin can be applied to it. 
 Zhao et al. \cite{zhao2016toward} propose an indexing scheme that leverages Zero-hop distributed hash table (ZHT) to provide a scalable dynamic indexing service that is able to handle churns.
% \ssnote{@Shashwat and Suman please see if we can add some more analysis here.}
 Contrastingly, the system design presented in this paper is based on a simple yet effective lightweight dynamic partitioning approach. 
}
\begin{figure}[t]
\centering
\includegraphics[width=0.9\columnwidth]{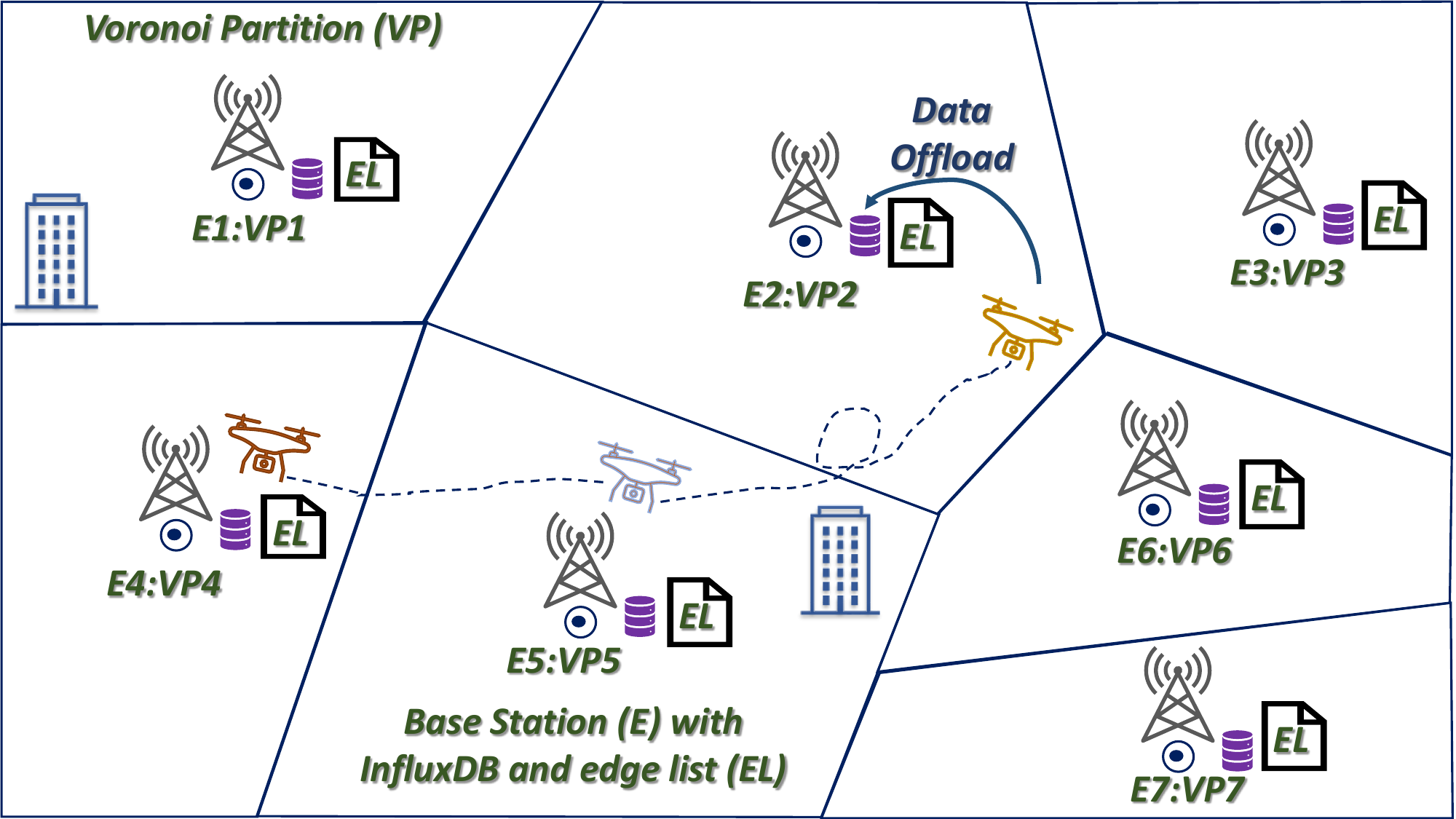}
\caption{\tdblite system design}
% \ysnote{YS to verify: Illustrate concepts from System model section (\S~\ref{sec:arch:system}), and Membership/discovery section (\S\ref{sec:arch:membership}) : drone, edge, voronoi partition, parent fog, data collection and offload to parent, influxDB on edge, celltower for edge, list of other edges with ID, polygon, NW endpoint, etc.}
% \srnote{Done}
\label{fig:arch:design}
% \includegraphics[width=1\columnwidth]{Motivation.pdf}
% \caption{Motivation}
% ~\\
% \label{fig-motivation}
% ~\\
% \includegraphics[width=1\columnwidth]{parent.png}
% \caption{Fog Partitions and Parent Allocation}
\vspace{-0.1in}
\end{figure}
\section{\tdblite Architecture}\label{sec:design} 
% \Note{4pgs}

% \textbf{SYSTEM MODEL}\\
% data sourced from mobile devices\\
% wide-area distributed edge servers\\

% \textbf{INSERTION}\\
% content-based replica placement\\
% indexing of shards\\
% vornoi partitioning\\
% onto InfluxDB services running on the edge\\

% \textbf{QUERYING}\\
% spatio-temporal queries using InfluxDB\\
% decentralized query execution\\
% load-balancing of query execution\\
% locality-aware query execution (client sending it to the "local" fog, and fog trying to exec locally)\\
% graceful degradation in performance\\

%A federated spatio-temporal database (§ ??) that uses lightweight content-based replica placement of shards (§ ??) for data sourced from mobile-edge devices (drone) onto wide-area distributed servers (fogs).

%% #####################################################
\subsection{\modc{Problem Scenario}}\label{sec:arch:system}
% \ysnote{SS to review}
% \ssnote{Finished pass. YS  may please take over}

We consider a multi-UAV environment where 100s of drones flying within the disaster-affected region in a city (or village), and continuously collecting sensor data using on-board sensors. They have limited on-board computing resources, such as a Raspberry Pi or Jetson Nano, to periodically (e.g., every few minutes) batch the collected spatio-temporal tuples into \textit{shards} and offload them for persistence. They may perform some initial pre-processing to record these as labelled time-series records, e.g., running a light-weight image classification model over video frames.

\modc{Edge servers are part of an on-demand Infrastructure as a Service (IaaS) offering, spread across the city. When provided by telecom providers, these servers are typically co-located with cellular base stations, whereas those offered by cloud or CDN providers are positioned within a few network hops of cellular towers. Each edge server has computational capabilities comparable to a low-end workstation or server, equipped with on-board CPUs and disk storage. These edge servers are interconnected over a dedicated network, enabling them to communicate, replicate data, and handle distributed query execution. In contrast, drones act as mobile data collectors that establish communication with the nearest base station for data offloading. They can interface with the co-located or nearby edge server to store data and execute queries, but cannot directly participate in the inter-edge communication network. Additionally, drones can also use their connected base station hosting the edge server to relay data to cloud resources. As drones move, communication handover occurs from one base station (edge server) to another, though the specifics of this mechanism are outside the scope of our work. While edge servers provide durable storage and computation, they are not guaranteed to have 100\% uptime and may experience transient failures, particularly in disaster scenarios—similar to cloud VMs.}

\subsection{Solution Approach}\label{sec:arch:approach}
% \ysnote{SS to review}
% \ssnote{Finished pass. YS  may please take over}

% We consider typical use cases where geospatial timeseries data needs to be processed in an environment which lacks  accessibility to an in-house cluster of servers hosted in a datacenter or a  public cloud. 
% To serve those use cases in particular, 
We describe the high level design of \tdblite here, and drill down into specific details in later subsections.

\tdblite follows a peer-to-peer decentralized storage approach to process spatio-temporal data across a collection of edge \modc{servers} that are provisioned across the MAN for hosting the distributed data store (\S~\ref{sec:arch:membership}). \addc{Unlike a fully distributed P2P system, where all nodes have equal roles, AerialDB's design leverages edge servers as more capable peers that handle data storage, query execution, and coordination, while drones interact dynamically with the nearest available edge server.}
Each edge \modc{servers} exposes a \textit{public service endpoint} to accept off-loaded data shards from drones that connect to them (\S~\ref{sec:arch:insert}). Each \textit{shard} has a unique \textit{shardID} and contains spatio-temporal tuples containing observations from the drone. 
  We create three \textit{replicas} of each shard, and use \textit{content-based hashing}, over their shardID, and the spatial and temporal mid-points of their tuple range, to determine the edge \modc{servers} to place them on (\S~\ref{sec:arch:insert:hash}). 
The coordinating edge forwards the shard to the replicating edges through a private service endpoint, which then insert the tuples in these shards into a local InfluxDB instance they run (\S~\ref{sec:arch:insert:influx}).
We also \textit{index} the spatial and temporal ranges of the shard tuples and their shardID, and replicate these index entries onto a subset of the edge \modc{servers} using \textit{content-based addressing} over the entire range (\S~\ref{sec:arch:insert:index}). The shardID and replica locations are recorded as part of the index for enabling near real-time access to a replica of the inserted shard directly from the spatio-temporal filter in the query.

Each edge \modc{server} also exposes query interface as part of its \textit{public service endpoint} (\S~\ref{sec:arch:query}). Query clients, which are often within the MAN and may even be drones themselves, invoke this endpoint with an InfluxDB query which has spatial, temporal, and/or shardID predicates, along with any domain predicates. 
The edge uses the spatial query range, the temporal query range, and/or the shardID to identify the edge \modc{servers} hosting the relevant indices using the same content-based addressing approach, and forwards the query to their private endpoint (\S~\ref{sec:arch:query:index}). This returns a list of shard IDs and their replicas that overlap with the query predicate. 
The edge then intelligently identifies a suitable subset among the edges hosting the shard replicas and forwards the query to them (\S~\ref{sec:arch:query:plan}). 
The edges receiving the query execute it on the local InfluxDB and return the result-set tuples back to the coordinating edge. The coordinating edge appends the result tuples from all edges and sends them back to the client. Duplicate tuples are not generated or returned.

% (\S~\ref{sec:arch:query:influx})

\ysnoted{are we doing streaming returns, as soon as first fog responds? Are we logging it and can we report time to first response for locality plots?}

This is a fully federated design with no single point of failure, other than for initial bootstrapping. All edge \modc{servers} perform similar roles.
As described later, the system is resilient and durable to querying even with the failure of up to two edge \modc{servers} with some degraded performance but no data loss, and resilient to querying for three or more failures with graceful loss in data if all three replicas for matching shards were lost (\S~\ref{sec:arch:query:resilience}).

\subsection{Membership and Discovery}\label{sec:arch:membership}
% \ysnote{SS to review}

The edges that host \tdblite are defined at deployment time, and known to all other edges. Each is identified by a unique \textit{edgeID}, and has a \textit{spatial location} and \textit{service endpoint address}. The edge locations are static, e.g., a cellular tower location. We use the edge location to spatially partition the city region using Voronoi partitioning, as discussed in \S~\ref{sec:arch:insert:hash}. Each edge also has the \textit{polygon} identifying the partition it is responsible for.

% \Note{
The edges exchange heartbeat messages among themselves to verify the availability of other edges. \ysnoted{TODO add feature}
% }\ysnote{Do we do this? How do we use this info?}
Drones and query clients discover the list of edges an their metadata out-of-band -- by having it be loaded \textit{a priori}, querying one of the known edges or by querying a cloud bootstrapping service for this static list.
Once this initialization is done, the system operates in a fully decentralized model.

When a drone is flying, it connects to the nearest edge \modc{server} for its communication and offloading. This is called its \textit{parent edge} and is the edge whose Voronoi polygon over which the drone is currently flying. This guarantees that the parent edge is the most proximate. \addc{The frequency of parent edge assignment update is based on the heartbeat interval and the mobility of drones.  }

%% #####################################################
\subsection{Shard Placement and Indexing} \label{sec:arch:insert}

% \ysnote{SS to review}

\begin{figure}[htpb]
\vspace{-0.1in}
\centering
\includegraphics[width=1\columnwidth]{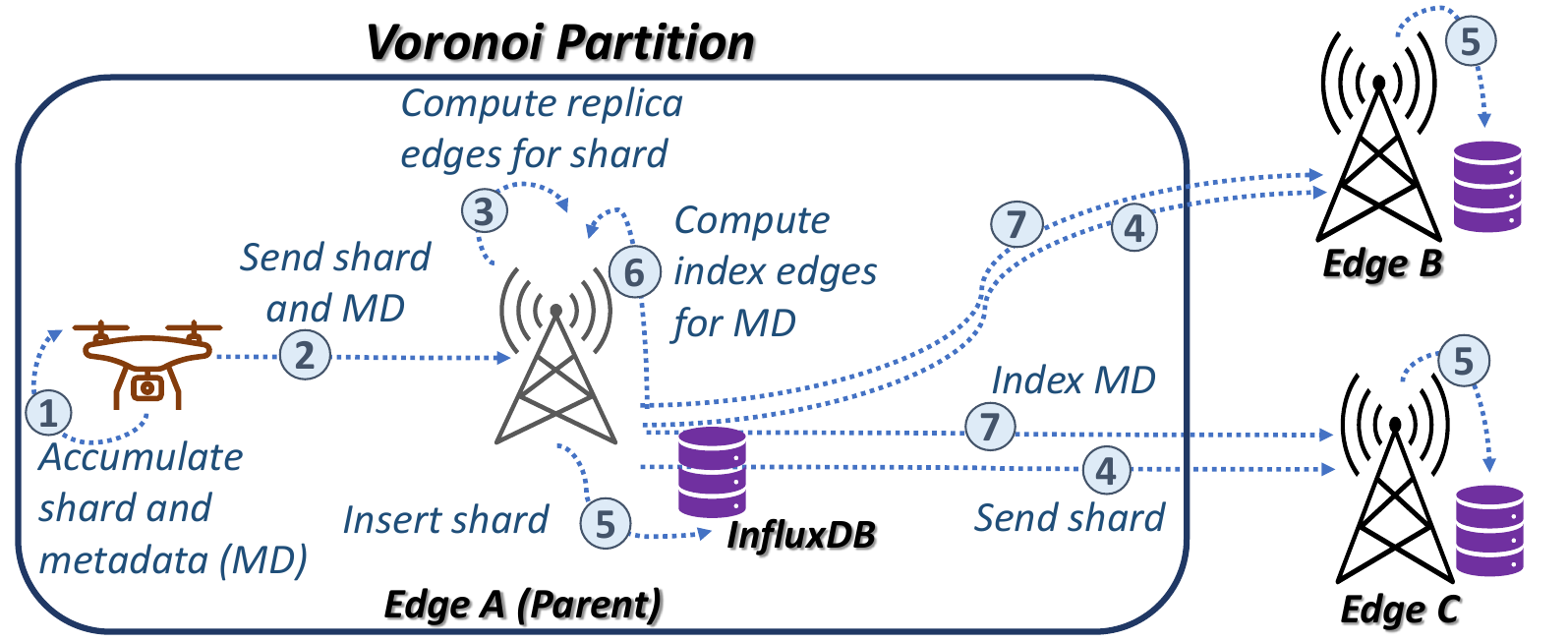}
% \includegraphics[width=1\columnwidth]{Fog only Insertion.pdf}
% % \label{fig-insertion}
% \includegraphics[width=1\columnwidth]{image1.png}
\caption{Execution flow for Insertion}
% https://lucid.app/lucidchart/e9936a99-318c-4794-badc-67ca214d7119/edit?viewport_loc=342%2C-533%2C1905%2C1095%2C0_0&invitationId=inv_e6f3c060-2786-419f-a681-e8d3cfafd3d8
\label{fig:arch:insert}
% \ysnote{@Suman: Corrections as per call...}
% \ysnote{@Suman: Merge these into single figure. Follow the description in \S~\ref{sec:arch:insert}.\\
% Make it wide rather than taller. Increase font sizes to be legible and more uniform.\srnote{Done}}
\vspace{-0.1in}
\end{figure}

% \begin{figure}
%     \centering
% \includegraphics[width=1\columnwidth]{insertion-seq.png}
% \caption{Sequence Diagram for Insertion}
% \label{fig:arch:insert-seq}
% % https://lucid.app/lucidchart/invitations/accept/inv_7474793b-f7c9-4db8-b65d-ef64cf3b213c
% \ysnote{Drop once it is merged into insertion figure}
% \end{figure}

% The queries performed on \tdblite  can be broadly classified as insert and fetch queries.  The architecture diagram of  \tdblite  with respect to the above two classes of queries  is illustrated in Figure   \ref{fig-arch}. The metadata for each object stored in  \tdblite  is replicated and indexed using the hash functions described in subsection \ref{sec-hash}. The sequence of steps performed internally by \tdblite  to process insertion and fetch queries is illustrated in  in the sequence diagrams in Figures \ref{fig-insertion-seq} and \ref{fig-fetch-seq}, respectively. 
 %\tdblite; has a default replication factor of 3.  

When a drone invokes the service endpoint of its parent edge to insert a shard, it contains the \textit{shard payload} with the spatio-temporal tuples, and \textit{metadata} that indicate the spatial bounding box range (latitude and longitude of top-left and bottom-right corners) covered by the tuples, their temporal range (start and end time stamp), and a globally unique shardID (e.g., a UUID). The parent edge coordinates the insertion process consisting of replicating the shards and indexing its metadata. This is described below and illustrated in Fig.~\ref{fig:arch:insert}.

\ysnoted{In future, we should send shards first to edges before indexing the metadata.}

%% =======================================================
\subsubsection{Content-based Hashing} \label{sec:arch:insert:hash}

\begin{figure}[htpb]
\centering
\includegraphics[width=0.8\columnwidth]{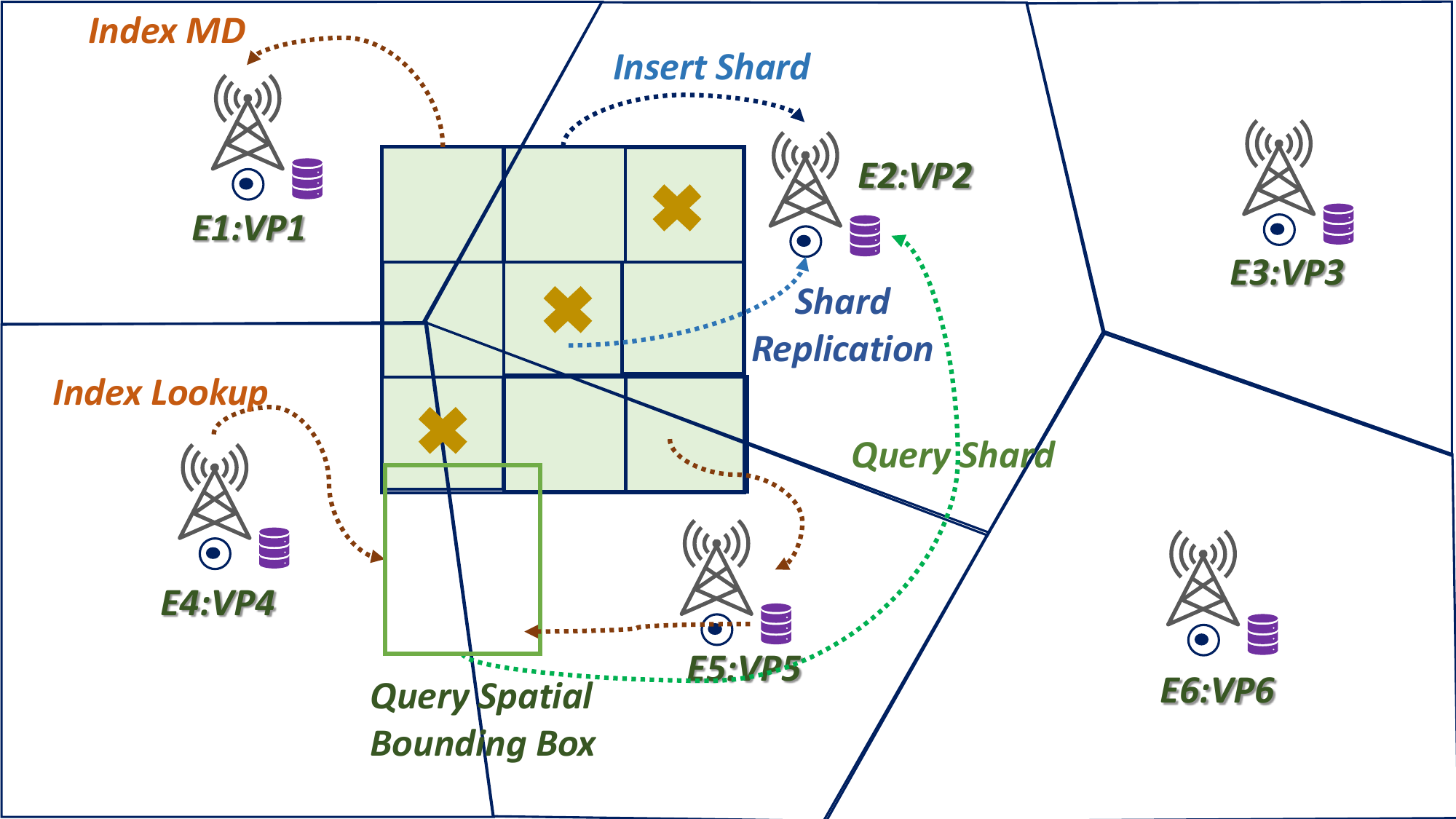}
\caption{Content-based hashing for indexing and shard replication}
\label{fig:arch:insert:hashing}
% \ysnote{@Suman: we need to update the figure to show the difference between gridded hashing to over-replicate indices and point-hashing to replicate the shards. Also show the challenges of the query not overlapping with the shard midpoint but with the shard range, allows the gridded index to help discover the shard.}
%
% \ysnote{verify that this is consistent with the experiments}\srnote{Will you provide a sketch?}
\end{figure}

\tdblite employs \textit{consistent content-based hashing}\footnote{\addcb{Following \cite{hansen2020content}, we have adopted content-based hashing  to handle the frequent occurrence of items that have not yet been seen by the users, i.e., for frequent occurrence of non-repetitive items...Prior papers \cite{broder2001using} have pointed out that having multiple hash functions tends to improve the performance by speeding up the look-up of indices...Especially, Voronoi clustering has been used in the past \cite{loi2013vlsh} for locality-sensitive hashing, and hence it is an obvious choice for spatial hashing in our case.}} 
% ~\cite{CAN}
to determine both the indexing location and the shard replica placement onto edges -- both at insertion and querying. This enables a decentralized design to process insert and query operations without any central repository for metadata.
 Prior papers \cite{broder2001using} have pointed out that having multiple hash functions tends to improve the performance by speeding up the look-up of indices.  
 Hence, we hash over three different properties: spatial, temporal and shardID, each with a different hash function:
\[ \mathcal{H}_s(lat, long) \rightarrow edgeID \]
\[ \mathcal{H}_t(timepoint) \rightarrow edgeID \]
\[ \mathcal{H}_i(shardID) \rightarrow edgeID \]
We sort the list of edgeID in ascending order and assign an index to each.
$\mathcal{H}_i$ is the simplest and just performs a modulo operation on the hash of the shardID with the number of edges, and the return value indicates index of the edge that the shard is placed in, $\mod(hash(shardID),~edgeCount)$. The hash function we use is \texttt{xxHash 64}, a fast non-cryptographic hash function. 
$\mathcal{H}_t$ is similar, except that it divides time into some fixed width range, e.g., $\tau=5~mins$, maps the timepoint into one of these range buckets, and takes a modulo on this bucketID to get the edge ID. E.g., if the timepoint were \texttt{2023-08-15 10:00}, this translates to a Unix epoch minute of X, a bucket ID of $Y = \big\lceil{\frac{X}{5~min}}\big\rceil$, and an edge index of $\mod(hash(Y),~edgeCount)$.
% \ysnote{@Shahswat: fix this}. 
Hashing the bucket ID ensures that periodicity in the collection time of the shard does not cause the same adjacent edges to be hit often. 

The spatial hashing is more involved, and uses \textit{Voronoi diagram}~\cite{doi:10.1142/9789814355858_0006}, which is frequently used to search for points or nearest neighbors in spatial and geometric queries. Especially, Voronoi clustering has been used in the past \cite{loi2013vlsh} for locality-sensitive hashing, and hence it is an obvious choice for spatial hashing in our case. A Voronoi diagram (or tessellation) partitions a 2D Euclidean space having $n$ points (called \textit{sites}) into $n$ convex polygons (\textit{cell}), one per site, such that all points within that polygon are closer to that site than any other site. 

Specifically, we construct a Voronoi diagram to partition the city into polygons (cells) for each edge (site), such that the edge is the closest to any point within its partition. Once the partitioning is done, the hash function $\mathcal{H}_s(lat, long)$ reduces to finding the edge within whose partition the hashed latitude and longitude falls within. We employ Fortune's algorithm~\cite{10.1007/BF01840357} to perform the partitioning.

These hash functions are consistent and static for the entire system, and known globally to all entities.

\addcb{\addcb{\paragraph{Discussion} 
Content-based hashing have an inherent load-balancing bottleneck when the content being inserted have the same values. This can surface in our system when there is temporal or spatial clustering of drones, i.e., when 100s of drones are simultaneously creating shards with the same timestamp, which gets hashed
% However, drones with synchronized clocks may hash shards 
to the same edge server using $\mathcal{H}_t$. While this may be manageable for smaller drones fleets ($\approx 100$s of UAVs) or if the shard generation rate is modest, it becomes more pronounced for large fleets (1000+ of drones) or high-frequency data, leading to potential hotspots on specific edge devices. This is a design trade-off caused by the need for decentralized and resilient spatio-temporal querying.
There are some design alternatives that can shift this overhead. E.g.,  we can alleviated this by batching data uploads or by introducing a small random delay ($\leq10$ seconds) before insertion, which helps distribute the load more evenly across edge servers but may affect the freshness of data.} Moreover, one could explore a trade-off between indexing and querying overheads, for example, by modifying $\mathcal{H}_t$ to yield $k$ unique edges instead of 1 for every time instant and indexing the shard randomly at any one out of those edges. This would alleviate the temporal load on a single edge server but increase the querying time since up to $k$ lookups are required. In our design, we are prioritizing fast, decentralized and resilient spatio-temporal querying over the potential temporal hotspots that can occur with specific workloads.}
\subsubsection{Shard Replication and Insertion into InfluxDB} \label{sec:arch:insert:influx}

% \par Using the index metadata information from the fogs, the insertion or fetch queries are processed at the index fogs, and the data (block) is stored on the InfluxDb instances hosted there. The data is replicated among multiple fogs based on the same hash functions. Similarly, the fogs where the replica of the data is to be stored, i.e., replica fogs,  are also determined using the same hash functions. Finally, \tdblite  executes the insert queries on the \tdblite  and InfluxDb instances hosted on the replica fogs to store the  metadata index and the replicas of the data.
% The insert queries are submitted by the user or the client application to an edge which chooses a coordinator fog for that operation as follows.  As illustrated in Figure \ref{fig-arch}, the proximal fog (i.e., Fog A) is determined using the Voronoi partitioning scheme described in Section \ref{sec:design}, and designated as the coordinator fog. The client prepares the metadata and sends the block to be inserted along with the metadata to the coordinator. 

The parent edge receiving the shard and its metadata applies the respective hash function on the mid-point of the spatial range of the shard, midpoint of its temporal range, and the shardID to get three edgeIDs on which to replicate the shard. \modc{In case the edgeIDs produced by the hash functions overlap, or if a specific edgeID is not available due to failure or inaccessibility, the corresponding replica is placed on the immediate successor edgeID in a deterministic sequence.}
\ysnoted{This will break if we do query routing based on content based hashing and that edge does not have the relevant shard. Will affect performance, not correctness.}

Such an approach of hashing over different dimensions has the dual benefits of replication to allow durable access to the shard even with edge failures, while also decentralized querying based on the content. This is in contrast to others who may use three different hash functions on the same dimension for the replicas~\cite{krishna2023using}, or just one hash function but use its two successors as well~\cite{decandia2007dynamo}.

The shard is then sent to the three distinct edges concurrently by the parent edge. Each inserts all the spatio-temporal tuples in the shard into their local InfluxDB service and acknowledges to the parent fog. 
% \ysnote{@Shashwat: add a few lines on InfludDB schema, indexing/field groupings, tables, use of shardID as special field, etc.}

% the various filters (i.e., geospatial, temporal, or block id) applied in the queries to determine the fogs (i.e., Fog A, Fog B, and Fog C) where the replica of the index metadata must be stored. The coordinator assigns a unique block id and sends the block and metadata to the parent fog. Subsequently, the parent fog computes three sets of hash functions $H_x$, $H_t$, and $H_z$, i.e., hashing based on geospatial, temporal, and block id, respectively, with respect to the geospatial filter, the temporal filter, and the block id, to determine three different replica fogs to store the replicas of the block. The hash functions  $H_x$ and $H_t$ determine the target fogs based on  the geospatial range and temporal range of the query, respectively, while  $H_z$ is based on the block id that is inserted. 

%% =======================================================
\subsubsection{Distributed Indexing using Slicing}\label{sec:arch:insert:index}
Since the shards themselves are distributed across different edges and the InfluxDB instance within them, we need a means to discover the relevant shards and edges holding them at query time. This allows us to localize the querying to specific edges rather than do a broadcast to all edges, which unnecessarily introduces compute and network overheads, and added latency.
The index itself has to be federated and not centralized on a single edge. At the same time, just placing the index on the same edge as the shard will not suffice. The spatial or temporal hashing used to decide the edge to replicate the shard is based on the mid-point of that property for the shard. The bounding-box spatial query or time-range temporal query may itself be arbitrarily large or small. It may not overlap with this mid-point but at the same time overlap with the full spatial or temporal range of the shard.
% \ysnote{do we need a figure to explain this?}

We address this in a unique manner by indexing this shard not just on the replica edges, but on all edges that overlap with its spatial and temporal bounding box.
Specifically, we perform a fixed size partitioning of the spatial range and the temporal range into \textit{slices}, hash each slice to an edge, and index this shard on all the edges that overlap with the spatial or temporal range of this shard. The hash functions used are the same as above, in \S~\ref{sec:arch:insert:hash}.

This means that the index entry will be present in at least the three edges holding the shard replicas but also in many more edges, depending on the spatial and temporal range covered by the shard.
The indexing request is sent to these edges concurrently by the parent edge.
While this may be a larger number, the index entry is small in size and should not pose a severe overhead.
% \ysnote{need to see if this is to blame for 4VM insertion overhead, dark green stack}

We maintain this index in-memory using three different data structures, one each for spatial, temporal and shard index. The index allows us to rapidly search for any shard whose spatial or temporal range overlaps with the query, return its shard ID and replica locations.

% \ysnote{the target of the index is shardID and the replica locations.}

Once all indexing is complete, the parent fog responds to the client that the insertion was successful.
Over-replicating the indexing on multiple edges allows for decentralized querying and supports any arbitrary spatial and temporal range predicate. Further, it also enhances the reliability of the system by surviving more edge failures to localize the shards. This has been illustrated in Fig.~\ref{fig:arch:insert:hashing}.

% \ysnote{having indices across three dimensions also allows us to lookup based on just one of them present in the user query}

% In this manner, the block metadata is sent to three replica fogs, determined by the three hash functions, to index the block being inserted. Additionally, the metadata is also sent to all other fogs which A) have their partitions overlapping with the spatial span of the block, and B) are mapped to the time chunks in the temporal range of the block. This metadata replication facilitates the block to be locally accessible from the target region and also eliminates the need for a central lookup service for supporting spatial and temporal queries in real-time. Subsequently, the ids of the above target fogs are appended to the metadata, and metadata is indexed at the  \tdblite  instances hosted in each of the replica fogs. The block is indexed in the \tdblite  instances in the fogs, and stored in the InfluxDB instances in these  fogs.

%% #####################################################
\subsection{Decentralized Querying} \label{sec:arch:query}
% \ysnote{SS to review}

\begin{figure}[htpb]
\vspace{-0.1in}
\centering
\includegraphics[width=.8\columnwidth]{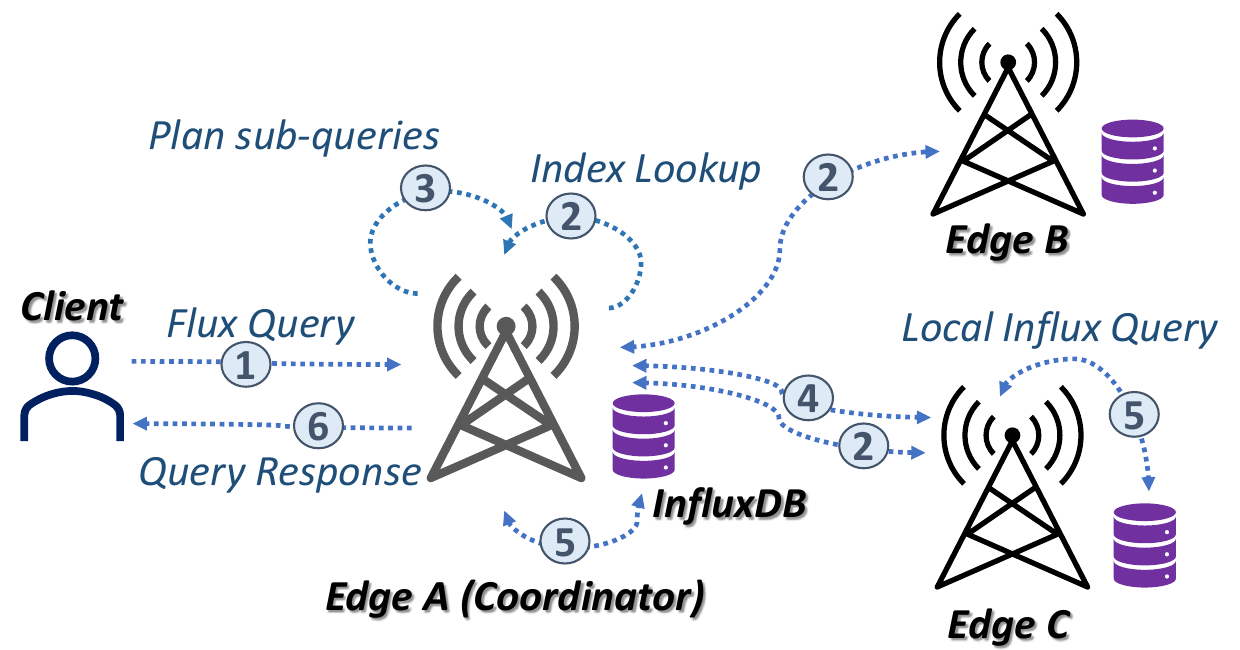}
% ~\\
% \includegraphics[width=1\columnwidth]{Fog only Querying.pdf}
% \caption{Fog only Querying}
% \label{fig-querying}
% ~\\
% \ysnote{keep only one of these three/merge?}
% \includegraphics[width=1\columnwidth]{image2.png}
% % \caption{Query Lifecycle}
% % https://lucid.app/lucidchart/e9936a99-318c-4794-badc-67ca214d7119/edit?viewport_loc=342%2C-533%2C1905%2C1095%2C0_0&invitationId=inv_e6f3c060-2786-419f-a681-e8d3cfafd3d8
% \label{fig-lifecycle}
% % https://lucid.app/lucidchart/f8457233-9f7f-4ff1-bd7c-fcf946967223/edit?viewport_loc=-4275%2C-1062%2C11026%2C5943%2C0_0&invitationId=inv_83956537-336a-4ebd-bad6-0993dcbf38ae
% ~\\
% \ysnote{do we need all three?}\\
% \includegraphics[width=1\columnwidth]{querying-seq.png}\ysnote{Drop once it is merged into query figure}
\caption{Execution flow of distributed querying}
% \ysnote{@Suman: Merge these into a single plot based on Sec 3.5 description.} \srnote{Done}}
% \label{fig-fetch-seq}
\label{fig:arch:query}
\vspace{-0.1in}
\end{figure}

Any client wishing to query \tdblite can invoke the query endpoint of any edge with an InfluxDB query to execute over the shards present in the overall system. The edge receiving the query serves as the \textit{coordinator} for this query. Broadcasting the query to all edges is inefficient, will not scale and can introduce duplicate results due to the replicas. Instead, we first use the index to identify the shards that may match the query, and then send the query to just one of the edges holding a replica of those shards. This is discussed next in detail, and the execution flow described in Fig.~\ref{fig:arch:query}.

% For any subsequent query, the query processing engine of \tdblite  performs a lookup on the corresponding Voronoi diagram whereby it searches  for overlaps of the Voronoi cells with the geospatial and temporal filter  supplied in the given query.  Based on the overlapping Voronoi cells obtained as a result of the above lookup, an edge (UAV) can find the concerned index fog for any query by performing localized search for index metadata within a pre-defined bounded rectangular patch determined by the boundaries of the aforementioned Voronoi cell. The above index fog is designated as a coordinator which is responsible for routing the subqueries to the replica fogs as elaborated later in Subsections \ref{sec:insert} and \ref{sec:query}. The coordinator applies hashing techniques (refer to Subsection \ref{sec-hash}) to determine the replica fogs where it routes the subqueries  for fetching (inserting) the metadata and the data from (to). Next, it  aggregates the result of the subqueries, and send the result to the client application or user.  

%% =======================================================
\subsubsection{Index Lookup} \label{sec:arch:query:index}
The query that is received must have a spatial, temporal and/or shardID filter to use the index; else it degenerates to a broadcast query. Most practical queries tend to be spatio-temporal in nature and meet this requirement.

If a spatial or temporal filter is present, we apply the same slicing technique described in \S~\ref{sec:arch:insert:index} on indexing, and partition the spatial bounding-box or the time range into slices, and hash these to retrieve the set of edges that have indexed the relevant shards. If a shardID is present in the query, a similar hashing returns the edge that has indexed this shardID. Specifically, for each of the query spatial, temporal and/or shardID predicates present, we get a corresponding edge set, $\mathbb{E}_s, \mathbb{E}_t$ and $\mathbb{E}_i$.

% \ysnote{@Shashwath: verify}
If these filters are combined using an \texttt{AND} Boolean expression, it suffices to lookup the index of just one of these edge sets, since that will return the relevant edges having shards that overlap with this filter. We select the set with the least number of edges to send the index lookup requests to: $\min_{E \in \{ \mathbb{E}_s, \mathbb{E}_t, \mathbb{E}_i \}}(|E|)$. This minimizes the number of edges explored and reduces the latency. Not all three filters may be present in a query, in which only the relevant ones are considered.
%
% \ysnote{@Shashwath: verify if we support this}
If the filters are combined using an \texttt{OR} Boolean expression, we need to lookup the union of all these edge sets, $E = \mathbb{E}_s \cup \mathbb{E}_t \cup \mathbb{E}_i$, since any one of them may match a filter.

The coordinator sends the lookup request to the relevant edges in parallel, and takes a union of the responses to get the set of shardIDs that are required to complete this query, along with the edges which hold their replicas. It is guaranteed that the query can be correctly satisfied by only examining these shards.

% As illustrated in Figure \ref{fig-fetch-seq}, the coordinator fog generates a shortlist of the possible target replica fogs containing the metadata index and the replicas of the data using the three hash functions  based on the temporal filter, spatial filter, and the block id supplied in the fetch query. The coordinator fog performs index lookup for information about the replica fogs where the requested blocks are stored in the metadata present in the \tdblite  instances hosted in the  fogs comprised in the above shortlist. 
   %The above fogs perform index lookup based on the above three hash functions and 

% As a result of the above lookup, the coordinator fog returns a list of block ids and the corresponding replica fogs hosting the said block ids.

%% =======================================================
\subsubsection{Query Planning} \label{sec:arch:query:plan}

In the query planning phase, the coordinator has a set of shardIDs and their replicas, $\{ sid, \langle e_i, e_j, e_k \rangle \}$, and it determines a ideal set of edge replicas to send the InfluxDB query to. A simple strategy us to randomly select one of the edge replicas at random for each shard to send the query to. This uniform random selection may have the serendipitous benefit of balancing the load across the edges. However, there are better guided ways of doing this to help reduce the execution latency, as discussed in the next section.

% "optimal" plan for the subqueries based on the cost estimates for each possible execution sequence for the said subqueries. 

% \ysnote{@Shashwath: Check query rewriting to include the shard IDs}
Once the replica edge for each shardID is selected, the coordinator edge creates an InfluxDB sub-query for each distinct edge it needs to query. This sub-query includes an additional filter that limits the query scope to just the shardIDs of interest on that edge's InfluxDB instance. This has two benefits: it reduces the search space for querying and speeds up the execution within the database, and it avoids duplicate results caused by the presence of another replica for a shard on the edge while that shard's replica is being queried for on a different edge.

\begin{figure}[htpb]
\centering
\includegraphics[width=0.9\columnwidth]{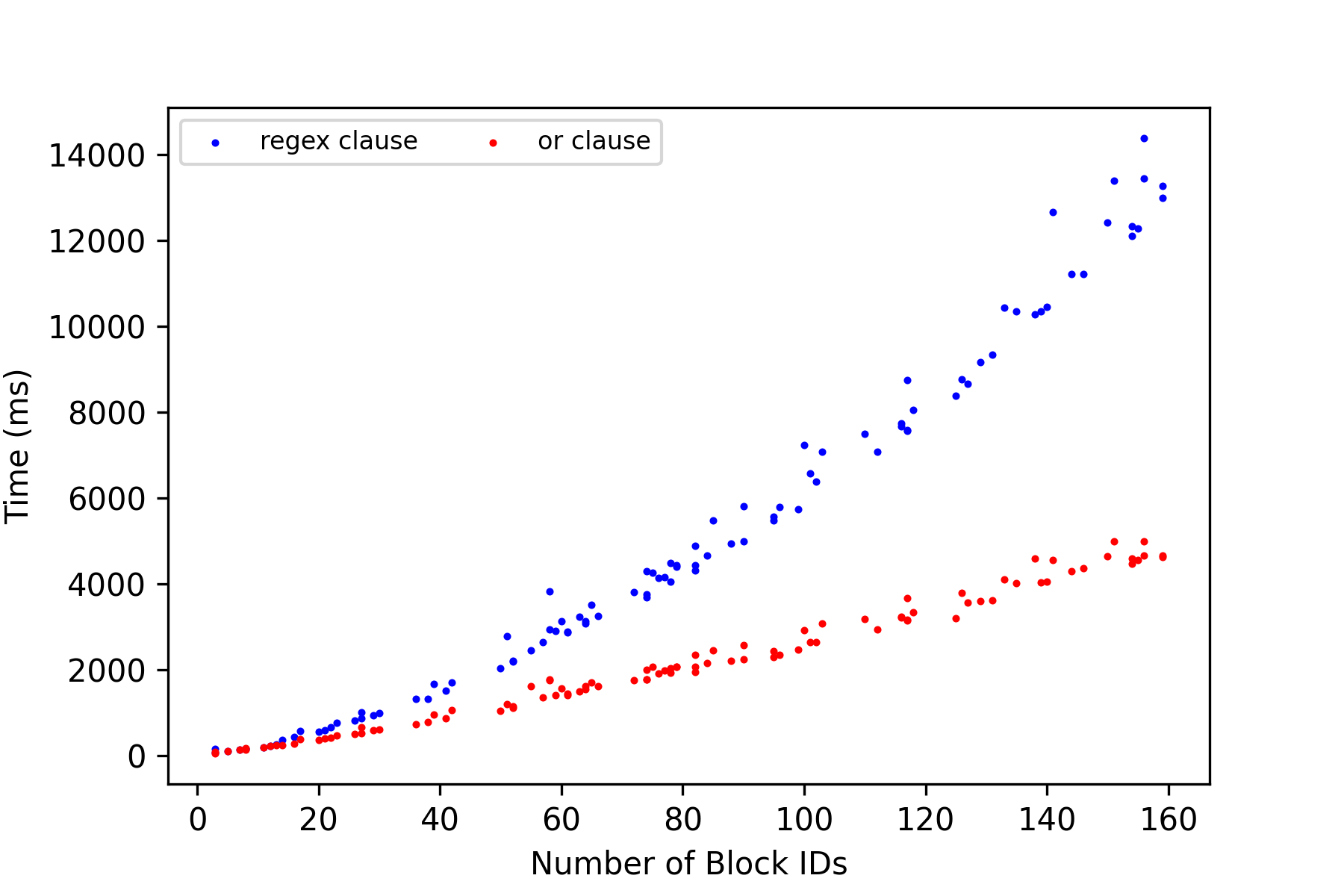}
\caption{\addcb{InfluxDB query latency for varying number of Shard IDs when expressed using OR vs Regex predicate.}}
\label{fig:arch:query:regex}
\end{figure}

These sub-queries are sent to the selected edges concurrently, where they are executed on that local InfluxDB instance. 
\modcb{We use InfluxDB's \texttt{OR} clause to match the shardIDs than a regular expression since it offered better performance compared to using regular expression matching. As illustrated by the microbenchmark in Figure ~\ref{fig:arch:query:regex}, the regex latencies (blue marker) grow exponentially with the number of blocks while it is linear and much smaller for the \texttt{OR} clause (red marker). Similar observations also also highlighted in some forums\footnote{\url{https://stackoverflow.com/questions/16638637/whats-faster-regex-or-string-operations}} and blogs\footnote{\url{https://www.deviaene.eu/articles/2019/using-regular-expressions-versus-string-operations/}}}
We enable strict spatial filtering on InfluxDB to ensure accurate spatial match.
If the list of shardIDs exceed 150, we then divide the sub-query into groups of 150 shards and execute them in parallel on the same InfluxDB instance to enhance performance.

\begin{figure}[htpb]
\vspace{-0.1in}
\centering
\includegraphics[width=1\columnwidth]{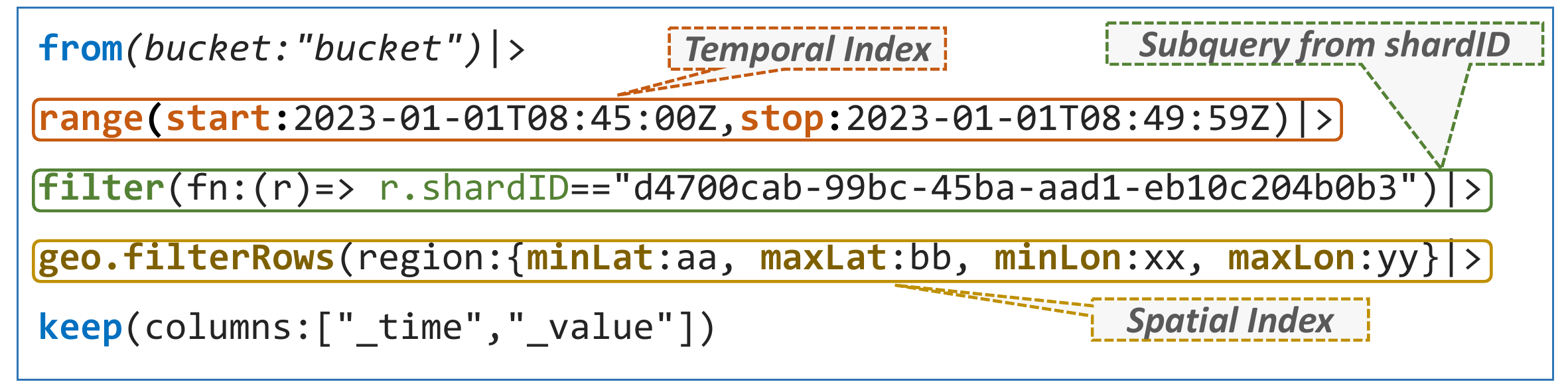}
\caption{Example Query. Spatial is anonymized.}
% \ysnote{@Suman: Verify that this is consistent with the experiments. E.g., do we support aggregates? Make it different looking than torqueDB to avoid revealing authorship. We may wish to mention "Index", "InfluxDB" and "Coordinator"}
% \ysnote{Lat/long should be masked}\srnote{Done}
\label{fig:arch:query:sample}
\vspace{-0.1in}
\end{figure}

% \ysnote{Anything else of interest here?}
The result-set tuples from these queries are returned to the coordinator edge, which streams the rows back to the client as they arrive. \ysnoted{TODO streaming response}
\modcb{As mentioned before, any aggregation operator present in the query must be scoped to a single shard for the results of the query execution on each edge to be returned independently.}
Figure~\ref{fig:arch:query:sample} illustrates a sample query.

% Subsequently, the chosen sequence of subqueries are executed on the \tdblite  instances hosted in the target replica fogs.  Next, \tdblite  executes the corresponding flux queries for the above subqueries on the InfluxDb instances in the replica fogs. On receiving the responses for the above flux queries, \tdblite  processes the responses of each subquery and sends these to the coordinator fog.   When responses for all the subqueries are received, the coordinator fog aggregates the results of these and sends the aggregated result to the client application or the user.

% \ysnote{@Suman/SS to discuss \ref{fig:arch:query:sample}}
% Figure~\ref{fig:arch:query:sample} illustrates how a sample query is currently executed with over a particular region in a city for a given time period. In the L3 stage of the query processing, the client applies the temporal and geospatial filters specified in the query, generates the subqueries, and chooses a replica fogs to fetch the metadata from and execute the subqueries on. Next in the L2 stage, the subqueries are executed on the replica fogs, and in the L1 stage, the results of the subqueries are aggregated. 

%% ----------------------------------------------
\paragraph{Load balancing shards}
Given $\{ sid, \langle e_i, e_j, e_k \rangle \}$, we need to select a subset of edges to ensure coverage of exactly one replica for each of the shards while attempting to minimize the execution time of all sub-queries sent to the selected edges.
There are two opposing intuitions that go beyond a uniform random choice and can improve the performance: (1) \textit{MinEdges}, which select the fewest number of edges, or (2) \textit{MinShards}, which select the most number of edges, such that they exactly cover one replica of all shards. 
\textit{MinEdges} attempts to reduce the number of distinct edges that are queries. This can have the benefit of making fewer network invocations from the coordinators and less static overheads for executing the sub-query on each edge. 
Alternatively, by selecting the most number of edges while just ensuring coverage, \textit{MinShards} tries to minimize the average number of shards that are queried on each edge as part of its sub-query. This can reduce the dynamic query execution time that increases with the number of shards queried, and maximizes the parallelism across edges.

\ysnoted{Another factor that affects the query time is the size of the edge database, but that is more or less uniform for all edges.}

% In our experimentation with varying blocks per query performed, we have observed a positive correlation between InfluxDB query latencies and the number of blocks queried in a single Flux Query. In other words, we have empirically seen that the latency increases with an increase in the blocks accessed per query.  To that end, \tdblite  have designed the load balancing algorithm $MinBlock$ as follows. The querying service receives the list of blocks that it needs to query for exhaustively answering the client query, from the index lookup. Subsequently, the load balancing module of \tdblite  distributes the blocks to be queried across the replicas available in an attempt to minimise the number of blocks queried per fog and maximise the number of fogs queried in total as follows. 

% \ysnote{YS to resume from here}

% \ysnoted{\shashwat{Is the cost-model algorithm required here?}\ssnote{@Shashwat may please update this subsection with the latest cost model.}}

The MinShards load balancing algorithm is described in Alg.~\ref{alg:loadbal:minshards}. Intuitively, we iteratively assign shard replicas to edges that have the fewest number of replicas. 
We first sort the edges by the number of replicas that they hold for this query. In ascending order, we select the edge with the fewest replicas and from it, select the shard which  has the fewest replica -- in the absence of failures, all shards will have 3 replicas and we end up selecting any random shard. We assign this shard replica to this edge, remove this shard from all edges, re-sort the edge list and repeat this.
Conversely, the MinEdges algorithm (not shown) sorts the edges in descending order of replica count, greedilly selects all shard replicas in the first edge, removes those shard replicas from the other edges, re-sorts the edge list, and repeats this process.
% Sort fog in ascending order of replica count\\
% Iterate through fogs in ascending order\\
% For each fog, select an unassigned shard replica on it such that it is present on the fewest number of fogs (this is usually 3 unless some fog has failed; so this is random but iterating thru fogs in ascending order.).\\
% Remove this shard from all fogs and update the fog's sorted order.\\
% Minfog is in descending order and add all replicas. Update replica count, re-sort, repeat.
\begin{algorithm}[htpb]
\footnotesize
\caption{Load balancing using \textit{MinShards}}
\label{alg:loadbal:minshards}
\begin{algorithmic}[1]
\Procedure{MinShards}{$ SList \langle sid, Eids[~] \rangle [~] \}$}
    
    \Comment \emph{Invert mapping from edge to list of replicas}
    \State $EList \langle eid, Sids[~] \rangle \gets$ \Call{Invert}{SList}
    \While{Unassigned Shard in $EList$}
        \State $AEList \gets$ \Call{SortAsc}{$EList$} \Comment\emph{Sort in $|Sids[~]|$ order}
        
        \State $minSid \gets $ \Call{MinReplicas}{$AEList[0].Sids[~]$} \Comment \emph{Get shard with least replica count from edge (0) having fewest replicas}
        \State $Map[AEList[0].eid] \gets minSid$
        \State $EList \gets$ \Call{RemoveReplicas}{$EList, minSid$}
    \EndWhile
    
	\State \Return $Map$
%%%%
\EndProcedure	
\end{algorithmic}
\end{algorithm}

\paragraph{Exploiting client mobility and edge locality}
% \ysnote{LC-0, LC-3}
% Similar to the insert queries, the client application submits a fetch query to the coordinator fog identified using the Voronoi diagram technique (refer to beginning of Section \ref{sec:design}). 
Query clients may be present in any part of the city, and sometimes even be mobile, e.g., if the drone itself queries over the shards.
While the client can send its query to any or its nearest edge coordinator, as an optimization, it can also use the content-based addressing over the spatial, temporal or shardID predicate to forward the query to the edge likely to already have several of the relevant shard(s) required to answer the query locally. We call this locality-aware coordinator (LC), as opposed to selecting a random edge as coordinator (RC). The use of consistent content-based hashing enables the client library to leverage this optimization.

This can be coupled with another locality-aware optimization which executes the InfluxDB query within the coordinator edge, if all shards required to answer the query are available locally (avoids additional network hop), and there are fewer than some $n$  number of shards to query over (avoids overwhelming the local edge). We identify this strategy as LC-$n$.

\ysnoted{\ysnote{does content based hashing come to play here? It does not! Maybe the client routing the request to the relevant local coord is an example? Takes into account the mobility? Later, clients could send the request to some temporally local coordinator?}\\
TODO add feature to start streaming results to client as soon as first fog responds}

\subsubsection{Query Resilience} \label{sec:arch:query:resilience}
Lastly, the query planning should survive the loss of one or more edges as well. Resilience is built into different layers of the query execution.

Firstly, clients can select any edge that is available for performing its query, which upon receiving the query, assumes the role of the coordinator edge for that query.
The edges exchange heartbeats to determine which other edges are currently available. The coordinator edge for a query can select the edge index set $\mathbb{E}_s, \mathbb{E}_t$ or $\mathbb{E}_i$ that does not have any edge failures, and among them, the one with the least number of edges. If none of the sets have an edge without a failure, then the index lookup has to be \textit{broadcast} to all edges, which can cause a degradation in performance. However, this lookup is lightweight and the latency penalty for the broadcast is small. This is guaranteed to be successful as long as we have 2 or fewer edge failures. Since the index may be over-replicated, it is possible that this may give a successful result even if 3 or more edges have failed, but this cannot be guaranteed.

From the list of edges that the index returns, the load balancing algorithm returns a set of available edges to send the sub-queries to. Here again, the loss of up to 2 edges may cause a marginal loss in performance but guarantee the correct result. Failures of 3 or more edges cannot guarantee a correct result.
Note that the InfluxDB query itself, which is costly, is not broadcast and executed on all edges in the presence of failures.

\section{Results}\label{sec:results}
% \Note{4.25pgs}
% \ysnote{YS has started reviewing...has write lock}

%% #####################################################
\subsection{Implementation} 
% \Note{0.5pgs}
The \tdblite \footnote{\url{https://github.com/shashwatj07/aerialdb}} distributed datastore is implemented using Java v21\footnote{\url{https://openjdk.org/projects/jdk/21/}}.
% which makes it lightweight as well as platform independent. This gives it a robust and secure runtime environment along with a stable ecosystem of extensible libraries and tools readily available in Java. %It is more lightweight than traditional distributed databases which allows it to be deployed on the edge, fog or cloud resources. 
% It uses Maven as the build automation tool, which helps simplify the process of building and packaging the software. %Maven is a popular tool in the Java development community, and provides a powerful set of features for managing dependencies, compiling code, running tests, and packaging artifacts. 
% By virtue of using Maven, \tdblite   is able to manage its dependencies more easily, and can quickly be built into a software ready for distribution.
%
It uses gRPC v1.53.0\footnote{\url{https://grpc.io/}} for remote execution of its services and Protocol Buffers (ProtoBuf) for data de/serialization.
% , potentially resulting in significant performance improvements over other serialization formats, such as JSON. Protocol Buffers are more compact and faster to encode and decode, which reduces the amount of network bandwidth required and improves the overall latency of the system.
%
We use the popular InfluxDB v2.6.0\footnote{\url{https://www.influxdata.com/get-influxdb/}} as the native spatio-temporal database that facilitates efficient storage and querying of time-series data using the declarative flux query language. That said, the interface dependencies InfluxDB itself is narrowly scoped and this can be replaced by an alternative like IoTDB or PostgreSQL easily.
%
% , enabling \tdblite   to handle high volumes of distributed system data in real-time and providing scalability, flexibility, compression, retention policies, and the flux query language which is highly declarative and functional for improved performance and ease of use.
%
We leverage Google's S2 Geometry library v2.0.0 \footnote{\url{http://s2geometry.io/}}
% \footnote{S2 Geometry is a library that handles spatial indexing and querying at a massive scale.} which provides a powerful set of APIs for working with 
for the in-memory spatial indexing, along with the JTS Topology Suite v1.19.0\footnote{\url{https://locationtech.github.io/jts/}} for spatial operations over geographic data.
% and performing spatial operations on it. 
Specifically, JTS supports Voronoi partitioning that is used by \tdblite.

Insertion and query clients running on the drone and other devices can use the gRPC interface in any suitable programming language to invoke the \tdblite endpoints on the edge \modc{servers}. We provide a default Java implementation.

\subsection{Experiment Setup}
% \ysnote{@Suman: Update this sub-section, talk about two deployment setups. YS to review after that.} \srnote{Done}

\ysnoted{
\begin{figure}[t]
\centering
\caption{\ysnote{[Low Priority] Heatmap of the distribution of dataset spatially among the WGS grids? Alongside the drone route?}}
\label{fig:result}
\end{figure}
\begin{figure}[t]
\caption{\ysnote{[Low Priority] Violin plot of distribution of query resultset size for 9 queries?}}
\label{fig:result}
\end{figure}
}

% \begin{figure}[t]
% \vspace{-0.1in}
% \centering
% \includegraphics[width=.8\columnwidth]{figures/10.png}
% \caption{Voronoi partitioning for 80 edges in the 100D setup. Lat/longs are anonymized. \ysnote{[TODO4] mask the lat/long and add as PDF}}
% \ysnote{Drop?!!}
% \label{fig:result:vornoi}
% \vspace{-0.1in}
% \end{figure}

% We use a 120-node IoT cluster configured and managed using the Internet of Things (IoT) emulation environment called VIoLET \cite{Badiger_2018} comprising 100 edge devices (0.1 vCPU, 100 MB RAM) that model drones and 20 edge devices (1 vCPU, 1 GB RAM) that model base stations. The drones are connected to the fogs over hierarchical 1Gbps switches with an average latency of 0.8ms. The containers run on a host server having an Intel Xeon 6208U CPU@2.9 GHz with 30 vCPUs and 128 GB RAM, running Ubuntu v18.04.

Our experiments use a 120-node IoT cluster configured and managed using the Internet of Things (IoT) emulation environment comprising 100 drone devices (0.1 vCPU, 100 MB RAM) that model drones and 20 edge devices (1 vCPU, 1 GB RAM) that model base stations. The containers run on a host server having an Intel Xeon 6208U CPU@2.9 GHz with 30 vCPUs and 128 GB RAM, running Ubuntu v18.04. In the setup, the drones are connected to the edges over hierarchical 100Mbps switches with an average latency of 10ms. \modcb{This closely emulates the network statistics that we would observe in cellular networks \cite{speed_claims, bandwidthplace2023, 4glatencypaper}.} The edge devices are connected to each other over a 1Gbps switch with an average latency of 1ms. 
For the weak scaling experiments, we use a 480-node setup with $400$~drones and $80$~edge devices where we replicate the $120$~node setup on $4$~host machines. 

For cloud experiments, we deploy a container with 16vCPUs with 64 GB RAM to emulate a high-end computing device on the cloud which runs on another host server. We run drone containers running on a separate host machine with 12vCPUs and 64GB RAM that connect directly to the cloud over the $100Mbps$ switch for data offloading.

\addc{The only exception to the above configuration is the distributed \tdblite and Feather implementation used in Figure \ref{fig:baseline-insert} and \ref{fig:baseline-query}, for which we use 1vCPU and 1 GB RAM for every drone device and keep everything else (fogs, network etc) the same. This is done to make the drones powerful enough to support a local data-store instance as required by Feather.}

% \srnote{This paragraph needs to be reviewed by Shashwat.}
% \tdblite, implemented in Java v11, runs on a 15-node cluster, configured with a uniform replication factor of 3. InfluxDB v1.7.9 runs on the fogs alongside \tdblite.

\subsection{Baselines}
\addc{We evaluate \textbf{\tdblite} against \textbf{Feather} \cite{9355826}, that we implement on our setup using InfluxDB as the underlying database. We configure Feather on a fixed topology of drones and edges, where five drones are connected to each edge server. For insertion, in Feather, the drones simply insert the shards in their local InfluxDB instance. On the other hand, for querying, in the absence of any spatio-temporal indexing in Feather, the queries fall back to a broadcast to the edge servers. For critical evaluation, we assume the best-case execution flow in Feather, where the data has been pushed up to the edge servers by the drones at the time of the query. This ensures that the drones are never queried and that all queries can be fully answered by the edge servers. However, in reality, Feather's design allows for lazy propagation of data from drones to edges, where the drones may be queried if needed to ensure freshness of the result.}

\modc{Additionally, we also evaluate our system against the Cloud baselines running \textbf{InfluxDB} and \textbf{MobilityDB} \cite{10.1145/3406534}  deployed on comparable hardware resources. The former is a NoSQL datastore, which is also the underlying datastore of our \tdblite and Feather implementation, whereas the latter is based on PostgreSQL and PostGIS.}

\addcb{In summary, we compare \textbf{\tdblite} with the following systems:
\begin{enumerate}
    \item \textbf{Feather}: A distributed edge data store.
    \item \textbf{MobilityDB}: A cloud SQL database optimized for spatiotemporal data. 
    \item \textbf{InfluxDB}: A cloud NoSQL timeseries database with spatial support.
\end{enumerate}}

%% #####################################################
\subsection{Insertion}

\begin{figure}[htpb]
\vspace{-0.1in}
\centering
% \centerline{\includegraphics[width=1\columnwidth]{stack_insertion.png}}
\includegraphics[width=.8\columnwidth]{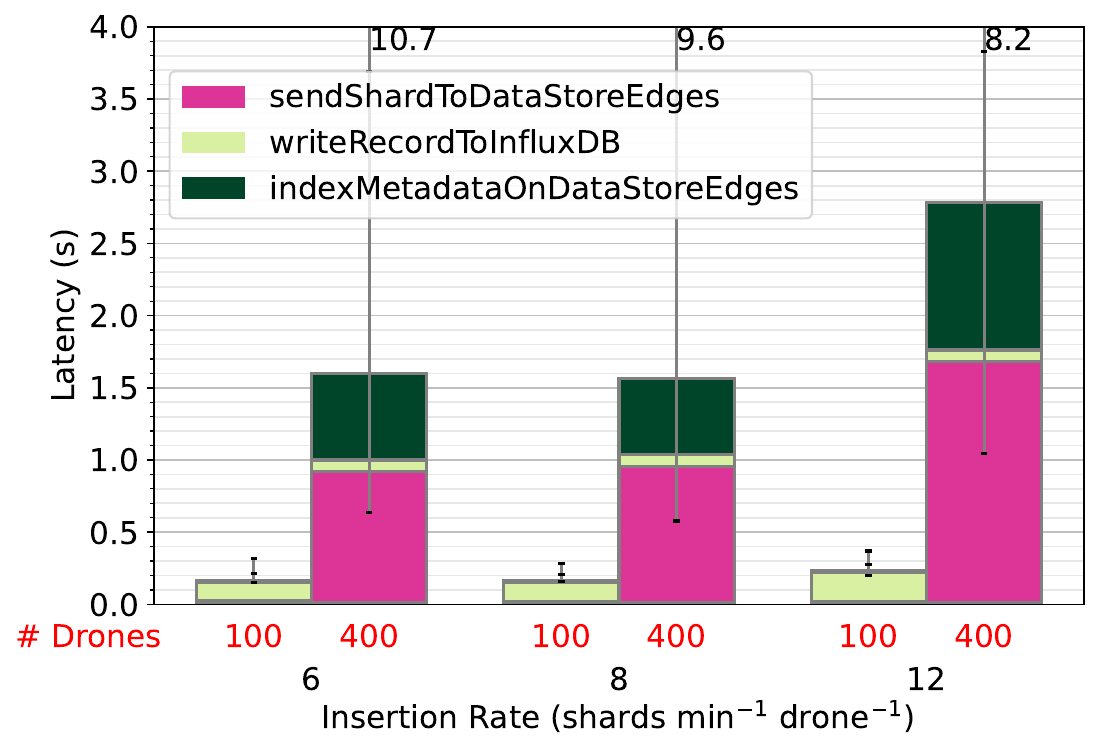}
\caption{Median \tdblite insertion latency for D100 and D400.}
% with different insertion rates}
\label{fig:result-insert}
\vspace{-0.1in}
\end{figure}
% \begin{figure}[t]
% \centering
% \caption{\ysnote{add figure 216 for RC vs. LC-0 vs. LC-3 locality}}
% \label{fig:result}
% \end{figure}

%% =======================================================
\subsubsection{Workload}
In our insertion experiments, we have two setups: one with 100 drones and 20 edge \modc{servers} in the city (\textbf{100D}), and the other with 400 drones and 80 edges (\textbf{400D}).
The drones collect one sample of data every 5 seconds and send a shard, comprising 60 records, for insertion every 5 minutes. These observations contain pollution and weather data collected from sensors present in the public domain\footnote{\url{http://datacanvas.org/sense-your-city/}}.
% \ysnote{What observations? Where was this sourced from?}
Altogether, we conduct these insertion experiments with 48 hours worth of data, effectively loading 576 shards per drone. Each shard is approximately 17~kB in size.
To simulate mobility in the drones, we use map data from OpenStreetMaps for \modc{part of Bangalore, India} that is around 20~km $\times$ 25~km in area. The drones perform a random walk along roads within the city with a velocity of 10~m/s. The trajectory visited by the drones match the spatial data present in the shard records.
We use cellular tower data from OpenCellID for the city and select 20/80 random tower locations to place the edge \modc{servers} at. \modc{We assume a practical $5-10m/s$ speed for all drones and ignore wind variability~\cite{9488740}. At every street intersection, the drone either hovers at the intersection $(P=0.8)$ or chooses one outgoing street $(P=0.2)$ out of the ones available at random. The
communication ranges (i.e., the distance of drones to parent edges) at any point in the experiment lifecycle
range between $1km$ and $5km$, the median being around $2km$. Attenuation and propagation are ignored for simplicity.}
%The Voronoi partitioning is done based on this. %, and shown in Fig.~\ref{fig:result:vornoi} for the 80 edge scenario.

% \ysnote{What is different for the 400 drone case?}

In addition, we also have a cloud setup with resources comparable to 16 edges, loosely the D100 setup, which runs a single instance of the InfluxDB/\modc{MobilityDB}. This forms the baseline for comparison with a centralized set-up.
% \begin{figure}
% \centering
% \begin{minipage}{.5\textwidth}
%   \centering
%   \includegraphics[width=\linewidth]{figures/insert_weak_scaling.pdf}
%   \captionof{figure}{A figure}
%   \label{fig:test1}
% \end{minipage}%
% \begin{minipage}{.5\textwidth}
%   \centering
%   \includegraphics[width=\linewidth]{figures/insert_baseline.pdf}
%   \captionof{figure}{Another figure}
%   \label{fig:test2}
% \end{minipage}
% \end{figure}

\begin{figure}[htpb]
\vspace{-0.1in}
\centering
% \centerline{\includegraphics[width=1\columnwidth]{stack_insertion.png}}
\includegraphics[width=.8\columnwidth]{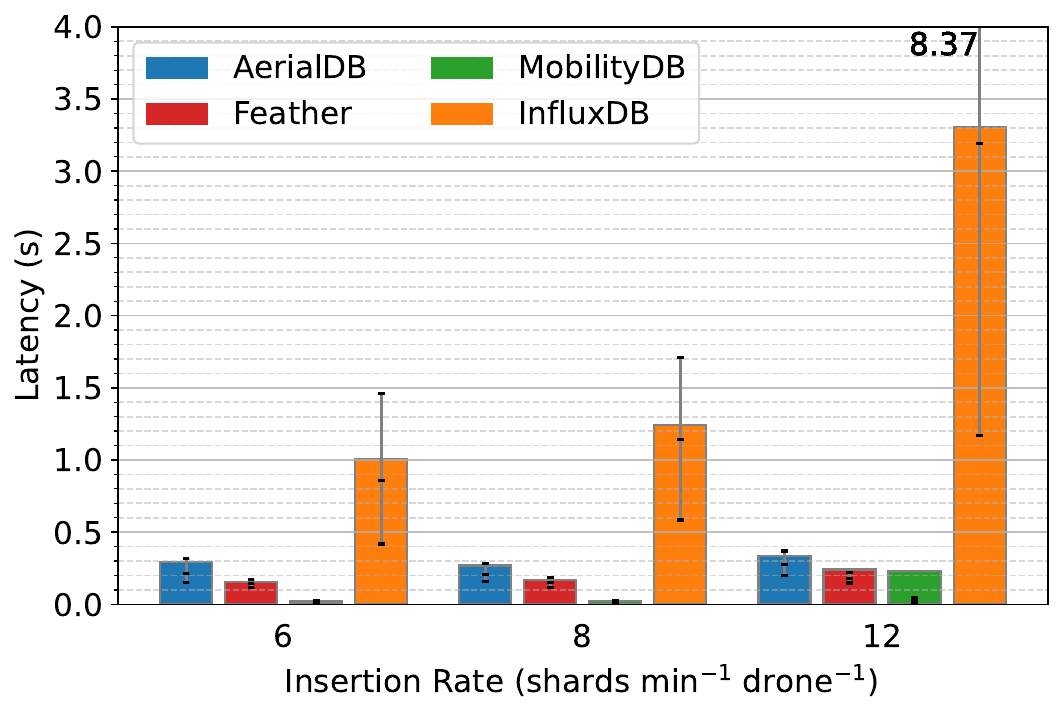}
\caption{Median, Q1 and Q3 latency for Insertion on D100.}
% with different insertion rates}
\label{fig:baseline-insert}
\vspace{-0.1in}
\end{figure}

\ysnoted{
\begin{figure}[htpb]
\centering
\includegraphics[width=1\columnwidth]{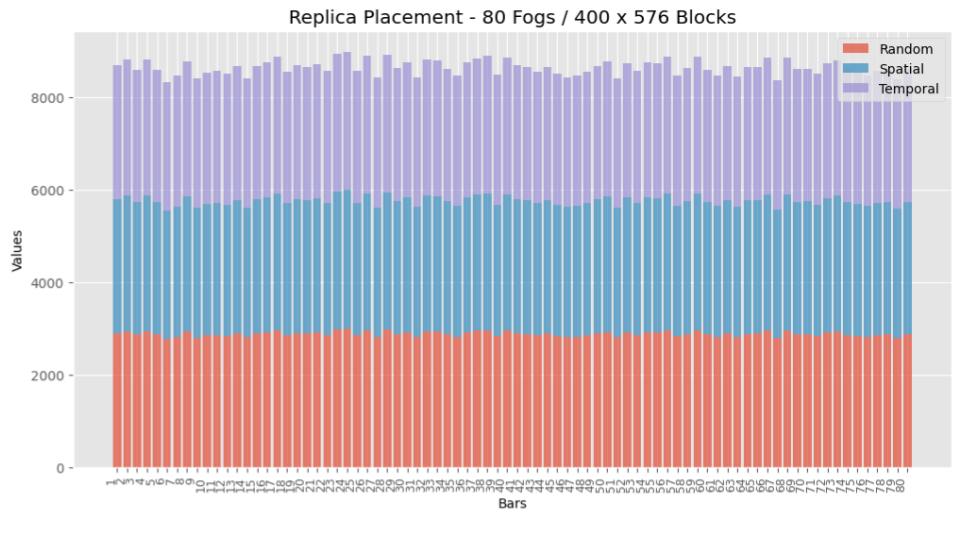}
\caption{Number of replicas placed on each edge for D400 using the spatial, temporal and shardID hashing.\\
\ysnote{[TODO3] ADD to the right Y axis, the number of index entries per fog per dimension as three marker. Change to PDF. crop top title.}}
\label{fig:result:shards}
\end{figure}
}
\begin{figure}[htpb]
\vspace{-0.1in}
\centering
% \centerline{\includegraphics[width=1\columnwidth]{stack_insertion.png}}
\includegraphics[width=0.85\columnwidth]{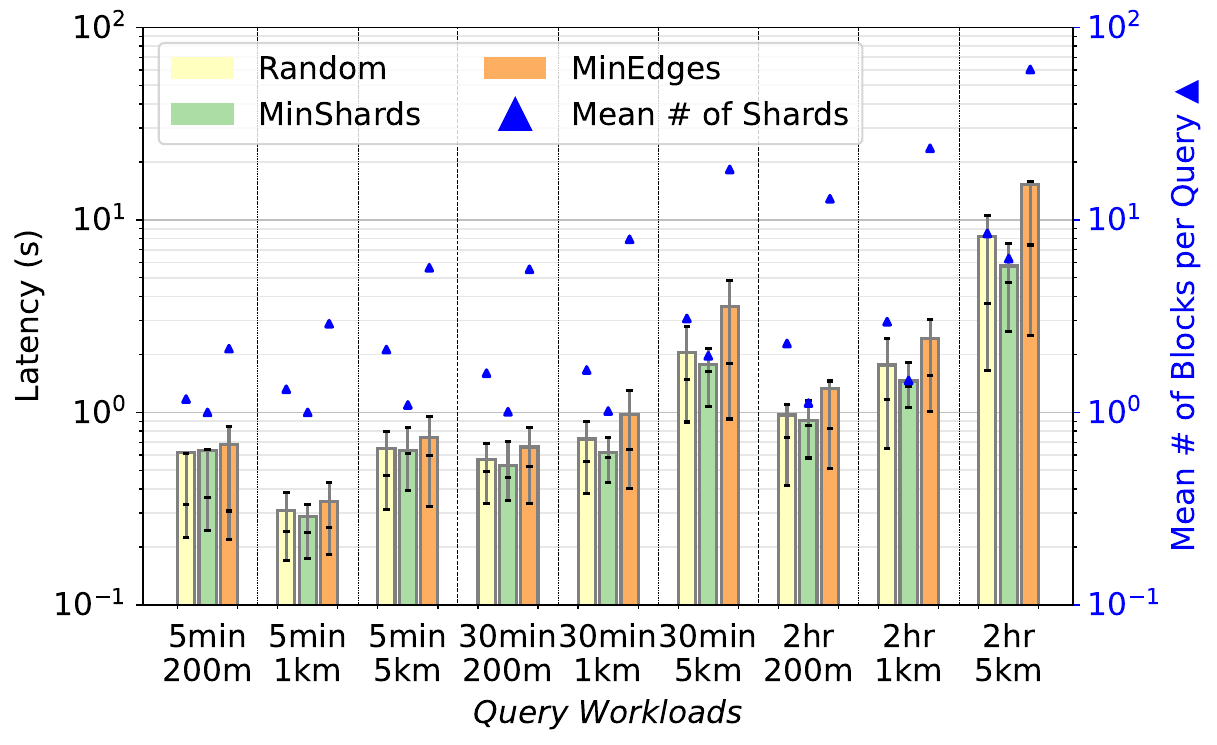}
\caption{100D Setup/16 concurrent clients}
% with different insertion rates}
\label{fig:cost-models-100}
\vspace{-0.1in}
\end{figure}

\begin{figure}[htpb]
\vspace{-0.1in}
\centering
% \centerline{\includegraphics[width=1\columnwidth]{stack_insertion.png}}
\includegraphics[width=.8\columnwidth]{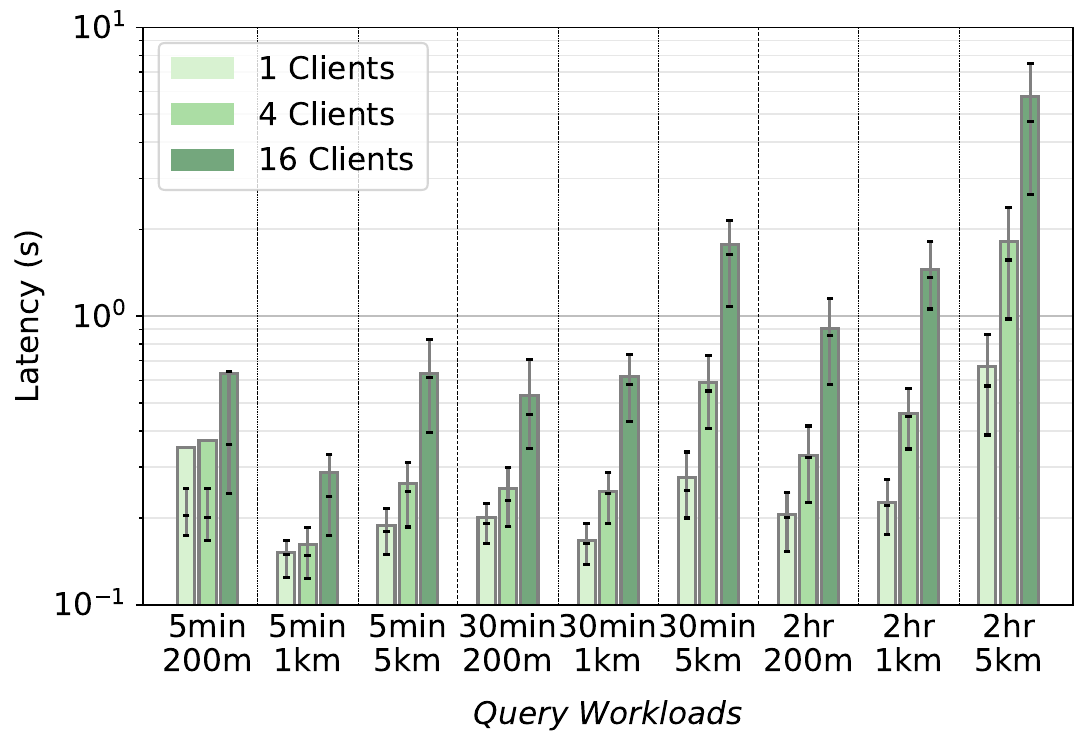}
\caption{100D Setup/MinShards}
% with different insertion rates}
\label{fig:scaling-query}
\vspace{-0.1in}
\end{figure}

\begin{figure}[htpb]
\vspace{-0.1in}
\centering
% \centerline{\includegraphics[width=1\columnwidth]{stack_insertion.png}}
\includegraphics[width=.8\columnwidth]{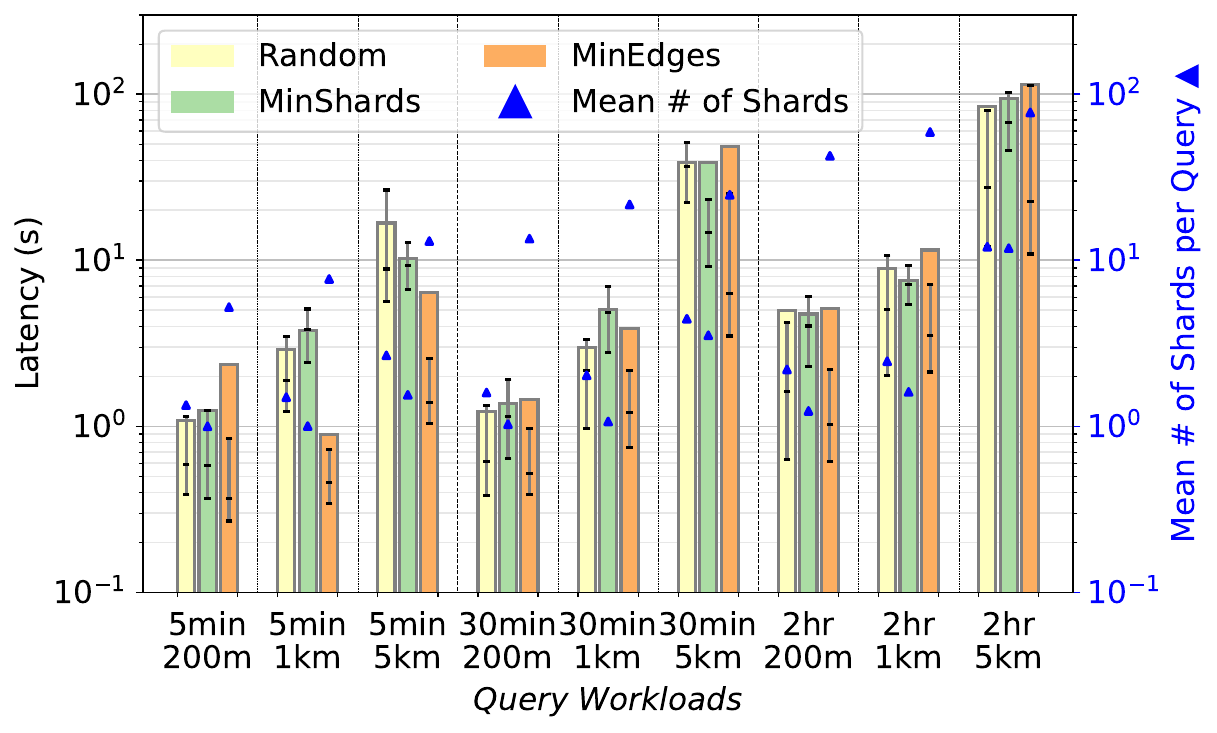}
\caption{400D Setup/64 concurrent clients}
% with different insertion rates}
\label{fig:result-models-400}
\vspace{-0.1in}
\end{figure}
\begin{comment}
\begin{figure*}[t]
\centering
\vspace{-0.1in}
\subfloat[100D Setup/16 concurrent clients]{
    \includegraphics[width=.33\textwidth]{5.pdf}
    \label{fig:cost-models-100}
  }
\subfloat[100D Setup/MinShards]{
    \includegraphics[width=.3\textwidth]{7.pdf}
    \label{fig:scaling-query}
% \caption{Effect of client concurrency on query performance (100D)}
      }
\subfloat[400D Setup/64 concurrent clients]{
    \includegraphics[width=.33\textwidth]{1.pdf}
    \label{fig:result-models-400}
  }
\vspace{-0.1in}\caption{Effect of load balancing and concurrency on query workload. Mean latencies are reported.}
% \ysnote{Triangles to be larger still in plot and smaller in legend.\\
% Vertical border bewteen 30min and 2h missing.}
\label{fig:result-models-scaling}
\vspace{-0.1in}
\end{figure*}
\end{comment}

%% =======================================================
\subsubsection{Scalability Analysis}
% disuss stack plot, where time is spent and why
% discuss the load balancing of replicas among the edges based on content based hashing
 The latency taken by the D100 and D400 setup for \tdblite when each drone is inserting at a rate of 6--12 shards/min per drone is shown in Figure~\ref{fig:result-insert}. 
% at different fast-forwarded controlled insertion rates, namely at rates of 1 shard per 5/10/30 seconds per drone, respectively.
% The fast forwarded rates are used specifically to stress the system to emulate the behavior of many real-world applications. On the other hand, we also observe sub-second scale  average query latency with realistic values as  space and time filters.
%
% 100 drones, 6 rate -> 0.216
% 100 drones, 8 rate -> 0.209
% 100 drones, 12 rate -> 0.275
% 400 drones, 6 rate -> 3.689
% 400 drones, 8 rate -> 3.463
% 400 drones, 12 rate -> 3.829

We see an increase in latency for the D400 case, taking $1.6$--$2.8~s$ per shard. Here, the time is dominated for sending the shard from the parent edge to the replica edges, and for the indexing of the metadata. We note that the 400 drone + 80 edge containers of the D400 setup run across 4 VMs on 4 different servers in our cluster. As a result, any service call between containers in different VMs incurs a higher network latency than the D100 setup where the containers are in a single VM. This contributes to the higher gRPC execution times for these two types of invocations.

% Figure~\ref{fig:result:shards} shows 
We also report that the number of shard replicas placed on each edge for the D400 setup is between 3846--4479, with the contribution from each of the hash being comparable. This indicates that we achieve a good degree of load balancing of the shards across all edges, and for all the hashes.

% %% ----------------------------------------------
% \paragraph{Scaling with insertion rate}

\subsubsection{Baseline Comparison}
\addcb{We focus on the D100 setup for a comparison with baselines as it is representative of the scale of mid-sized drone fleet deployments.
The mean, median and Q1-Q3 latencies of \tdblite D100 and all three baselines are reported in Figure \ref{fig:baseline-insert}. We see that the median insertion time for each shard for D100 is between $0.15$--$0.35~s$ for different insertion rates. This is much faster than the time taken by the single cloud instance having similar amount of resources as the cumulative number of edge servers. Here, a single instance of the InfluxDB is unable to scale its insertion with the number of concurrent drones, taking a much higher $0.9$--$3.2~s$ per shard. These queries are not natively parallelizable in InfluxDB, causing this performance bottleneck. \modc{Our design of using multiple InfluxDB instances, one on each edge, and explicitly performing parallel insertion scales much better and performs more than 10 times faster than the cloud.} Therefore, in addition to any network latency and cost penalties that may exist for the cloud, we also see a execution time performance drop relative to \tdblite.}

\addcb{Feather, on the other hand performs better than \tdblite due to no communication or data transfer involved in its insertion lifecycle, as it simply inserts data in the drones' local InfluxDB instance. Moreover, MobilityDB Cloud performs comparable to \tdblite at high insertion rates.}

%% ----------------------------------------------
% \paragraph{Scaling with number of drones and edges}

\ysnoted{
% Is number of duplicate index entries for 4VM much higher than 1VM to explain dark green bar? 
\textbf{Average number of fogs on which each shard is indexed, for 1VM (4.6 edges/shard) and 4VM (5.4 edges/shard).}\\
Maybe the higher gRPC time is to blame?}

%% =======================================================
% \subsubsection{Comparison with Cloud baseline}
% compare with Cloud baseline

%% #####################################################
\subsection{Querying} \label{sec:results}

%% =======================================================
\subsubsection{Workload}
We perform nine different query workloads, each varying in the temporal range ($5min,30min,2h$) and the spatial bounding box ($200m \times 200m, 1km \times 1km, 5km \times 5km$) queried upon. We have 200 queries of each type for D100 and 800 for D400. These are uniformly distributed among 1, 4, 8 and 16 concurrent clients for D100, and among 64 concurrent clients for D400. We also have a similar cloud setup with a single InfluxDB instance on a 16-core container, which we compare against our D100 setup.

%Random querying, Locality and Mobility, Reliability and Fault Tolerance

\begin{figure}[htpb]
% \vspace{-0.1in}
\centering
% \centerline{\includegraphics[width=1\columnwidth]{stack_insertion.png}}
\includegraphics[width=\columnwidth]{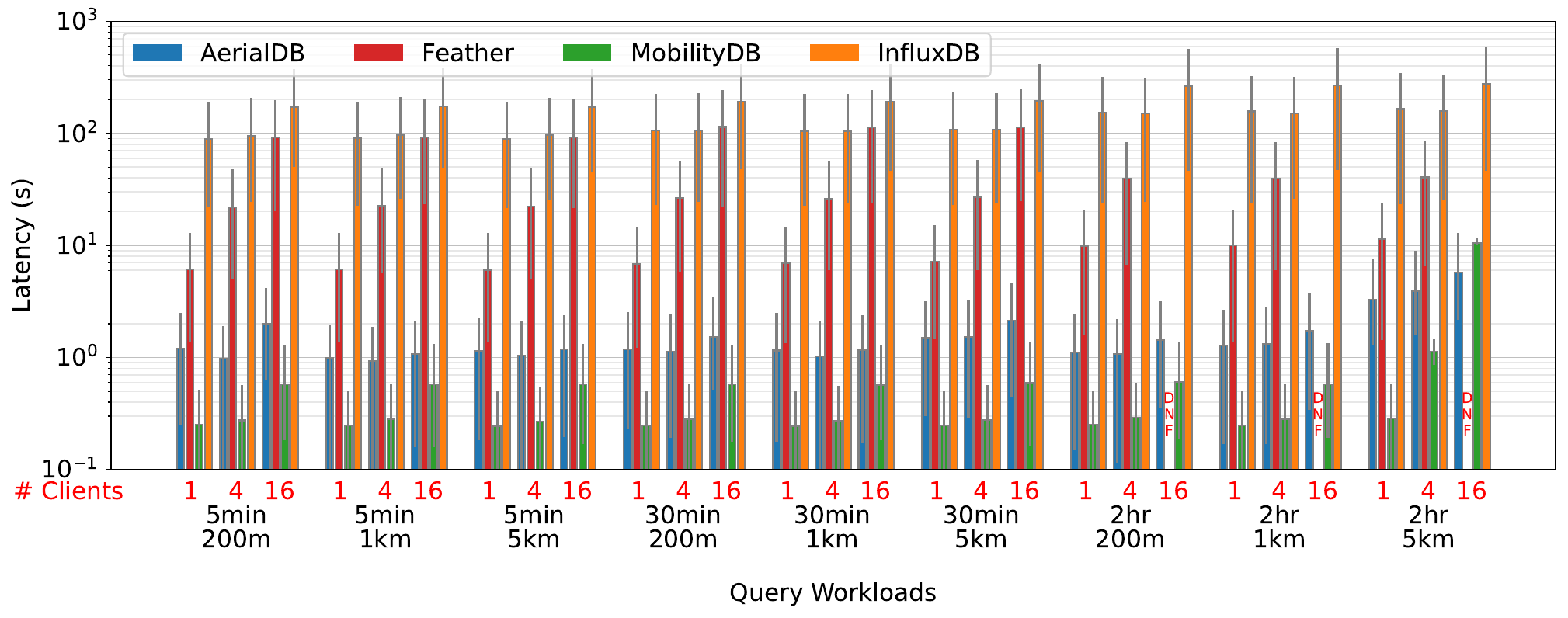}
\caption{Mean, Q1 and Q3 latency for Querying Workloads on D100. DNF implies Did Not Finish as queries timed out.}
% with different insertion rates}
\label{fig:baseline-query}
% \vspace{-0.1in}
\end{figure}

%% =======================================================
\subsubsection{Query Planning}

%% ----------------------------------------------
\paragraph{Effect of Cost Model}

% \begin{figure}[t]
% \centering

% \caption{Effect of load balancing on query performance ()}

% \ysnote{Make "Query Workloads" in 1pt larger font and italics. Make Y axes titles and labels 1 pt larger font. Rename as "Mean \# of Blocks per Query"\\
% Increase size of triangle marker by 2pts. Make boundry around bars 1pt thicker. Make whiskers 1pt thicker. Make Q1/Q2/Q3 1pt thicker and darker.\\
% I repeat: right and left Y axes have to have same order of magnitude range, in this case, 4-orders. Fix right Y axes to be from 10e2 to 10e6. Having grid lines that dont align with both left and right axes is meaningless.\\
% Have legend is 2x2 grid so that no marker overlaps with it.}
% \ysnote{Change all PNG to PDF}
% \end{figure}

Figure \ref{fig:cost-models-100} shows the performance of \tdblite for different query workloads, and when using different cost models. Left Y axis bars report the mean latencies for queries while the whiskers indicate the Q1--Q3 values. The right Y axis with blue triangle markers shows the mean number of shards queried, averaged across all the edges on which sub-queries are executed by the coordinator.

In general, we see that the mean execution time for the query workloads increase as the query's spatial and temporal range increases. The 5min/1km workload was the first to run, and hence exhibits a different behavior during the warm-up phase.

The relative performance of the three different load balancing algorithms, \textit{Random}, \textit{MinShards} and \textit{MinEdges}, are shown in the three bars for each workload. Each has their own merits for the reasons described earlier.
We see MinShards (green bar) performs marginally better in many cases, indicating that the load balancing across more edges offer better benefits compared to limiting the number of edges queried (orange bar). Even the random selection is occasionally competitive (yellow bar) due to its natural ability to spread the shards across edges.
The mean blocks per query (triangles) show the correlation between the strategy and the load on each query. As designed, the MinShards has the fewest shards queried on average for each of the query workload while MinEdges is the highest. This confirms the expected behavior.

% The $MinBlock-RC$ outperforms the other cost models while  $MinFog-RC$ seems to cost more in terms of the latency. 
We also report that the query time for the cloud setup (not shown) is \textit{$100\times$ slower} than the D100 setup for all the query workloads. This holds for 4--16 concurrent clients. This shows that a single-instance of InfluxDB running on the cloud, even with a large provisioning of resources, is unable to scale. The \tdblite query execution leverages the indexing that we have designed to limit the shards that are queried on InfluxDB and substantially reduces the query load even on low-end edge instances with just 1-core operating in parallel.

%% =======================================================
\paragraph{Scalability}
 % Figure \ref{fig:cost-models-100} demonstrate how different cost models, i.e., the shard placement strategies, perform for a 100 drone/20 fog setup. We instantiate 16 workers to collectively fire 200 queries, picking any fog at random. The $Minblock-RC$ strategy seems to outperform the others, followed by the $MinFog-RC$ strategy. 
%% ----------------------------------------------
Figure \ref{fig:scaling-query} shows the scaling of the D100 setup with 1--16 clients.
We see that having a concurrent client workload does reduce the performance of the system, but except for the largest query workload and some 16 concurrent clients cases, all others complete in under $1~s$.

Similarly, when we compare the scaling to D400 with 64 concurrent query clients (Fig~\ref{fig:result-models-400}), we see the latencies increase, partially due to the additional network time causes by the containers present across different VMs. Even here, the time taken per query ranges between 1--10~s.

% \paragraph{Scaling with size of query predicates}
% \begin{figure}[t]
% \centering
% \includegraphics[width=.8\columnwidth]{7.pdf}
% \caption{Effect of client concurrency on query performance (100D)}
% \label{fig:scaling-query}
% \end{figure}
% Figure \ref{fig:scaling-query} demonstrates that \tdblite  gracefully scales upto 16 clients.

%% ----------------------------------------------
% \paragraph{Scaling with concurrent query rate}

%% ----------------------------------------------
% \paragraph{Weak scaling with number of edges}

%  Figure \ref{fig:result-scaling-4VM} demonstrates that \tdblite  scales well with queries spanning over 80 edges.

% \ysnote{bring in NW latency of 4VM scenario as a reason for larger latency compared to 1 VM case.}

%% ----------------------------------------------
\paragraph{Effect of Locality and Mobility}
\begin{figure}[htpb]
\centering
\centerline{\includegraphics[width=.8\columnwidth]{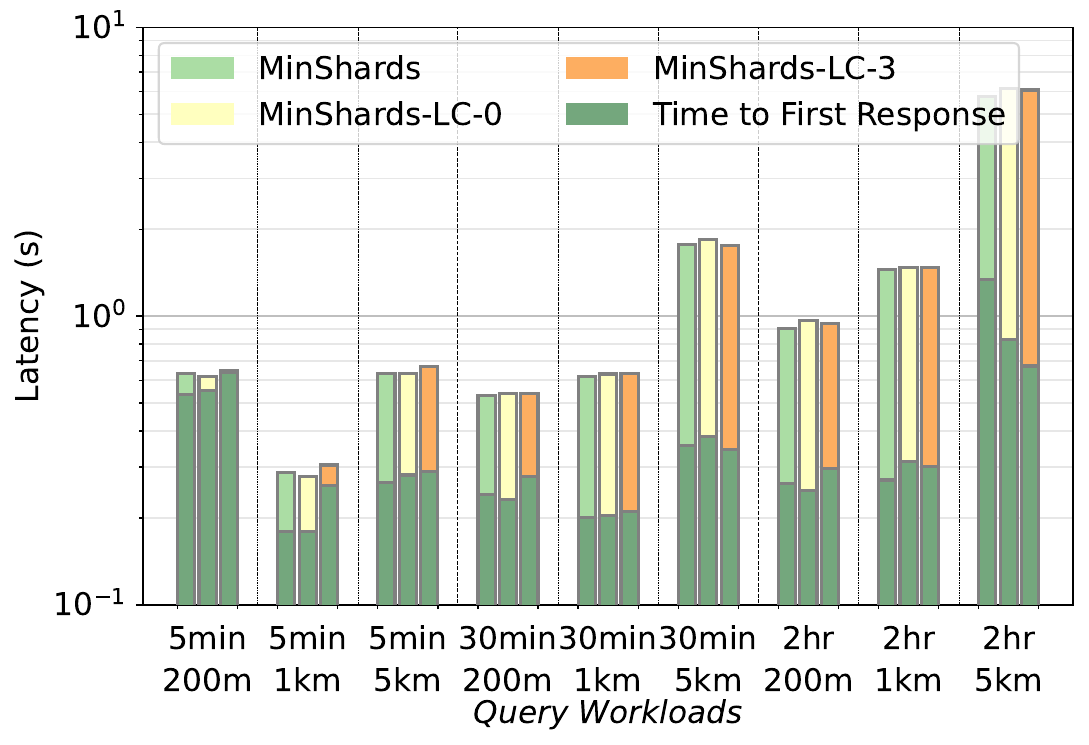}}
\caption{Effect of locality aware querying on D100}
\label{fig:result-RC}
% \ysnote{[TODO2]  are we doing streaming returns, as soon as first fog responds? Are we logging it and can we report time to first respone for locality plots}
\vspace{-0.1in}
\end{figure}

Figure \ref{fig:result-RC} demonstrates how \tdblite performs with different coordinator selection strategies, namely RC, LC-0, and LC-3. \addc{The default MinShards uses a random coordinator edge server to answer queries while MinShards-LC-$n$ uses the edge server situated closest to the centroid of the query's spatial predicate as coordinator, querying upto $n$ blocks locally (without involving any communication to other edge servers) whenever possible. One may expect that using the local coordinator reduces communication and may improve query performance. However,} all three perform similarly in most cases, partly reflecting the setup where the network latencies between the edge servers are low.
% As expected, the locality based coordinator selection strategies LC-0 and LC-3 outperforms the RC strategy in terms of the observed latency, where there is no apparent clear winner among the above locality based coordinator selection strategies. 

\subsubsection{Baseline Comparison}
\addc{Figure \ref{fig:baseline-query} shows the performance of \tdblite against the three baselines, i.e., InfluxDB cloud, Feather, and MobiltyDB . We observe that \tdblite exhibits a mean performance improvement of up to $97\times$ over Feather, and up to $185\times$ over InfluxDB Cloud, with bounded tail behavior. Moreover, it yields comparable performance with respect to MobilityDB Cloud for larger query workloads.}

\addcb{To statistically validate our performance claims, we first assessed the homogeneity of variances across different systems using Levene’s test, which confirmed that the variances were significantly different. As the data was normally distributed, we conducted Welch’s one-tailed t-tests at a 95\% confidence level to compare the performance of \tdblite against each of the baseline systems. These tests confirmed that \tdblite consistently outperforms all other systems across the evaluated workloads ($p\_values \leq 10^{-100}$ consistently), except for MobilityDB ($p\_values = 1.0$ in all except $(2h, 5km)$). However, even for MobilityDB, \tdblite demonstrates superior performance in the heaviest $(2h, 5km)$ workload ($p\_value = 0.0079$), as we have discussed above.}
%% =======================================================
\subsubsection{Resilience to edge failures}

\begin{figure}[htpb]
\centering
\includegraphics[width=1\columnwidth]{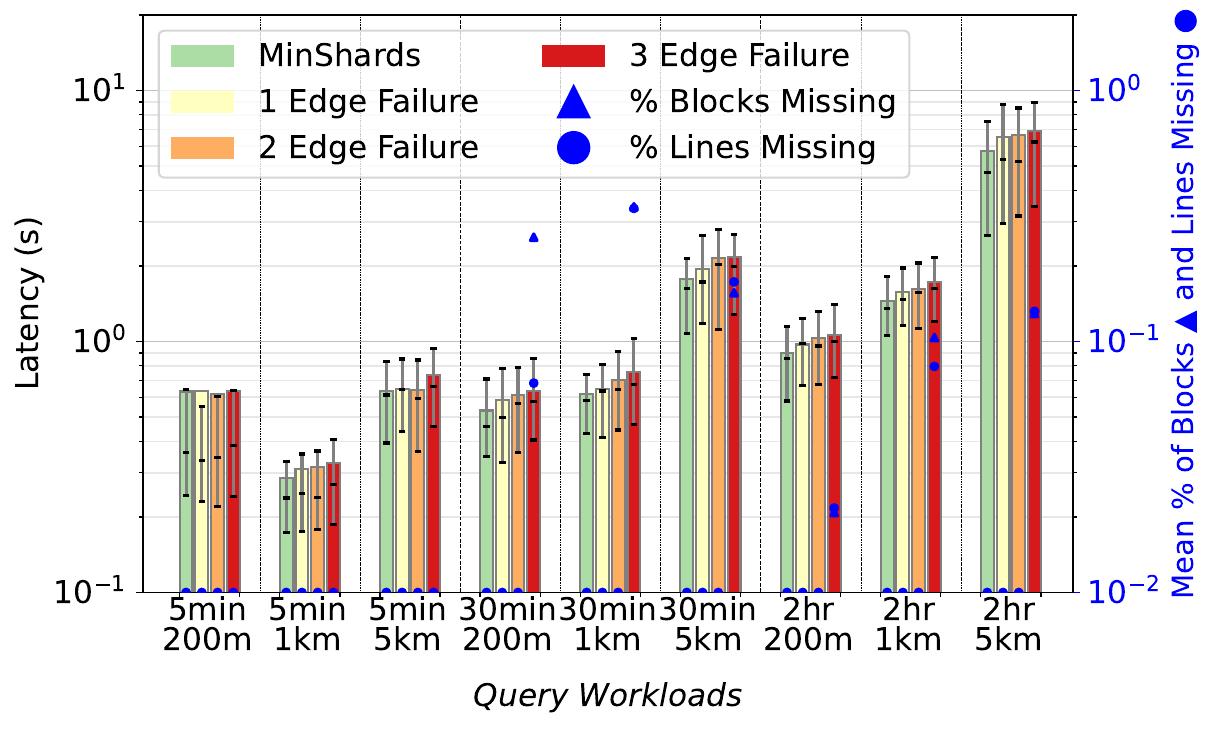}
\caption{Resilience of \tdblite querying when 0--4 edges fail in 100D (20 edge) setup.}
\label{fig:reliability}
\vspace{-0.1in}
\end{figure}
Figure \ref{fig:reliability} shows \modc{\tdblite D100} in the presence of edge failures. \modc{We simulate failures randomly, the probability of an edge failing being $1/20$.} We show that \tdblite can comfortably support up to 2 edge failures without compromising on performance across query types. Upon the 3rd failure, we start to observe up to 1\% loss in accuracy but without substantial increase in the observed latency. This confirms the resilient design of our framework.
% However, for the larger query workloads, we see that the LC-3 is able to get the initial result faster than the other strategies.

% %% =======================================================
% \subsubsection{Comparison with Cloud}

% \begin{figure}[t]
% \centering
% \includegraphics[width=1\columnwidth]{2.pdf}
% \caption{Comparison of \tdblite query performance with a cloud InfluxDB instance for 100D \ysnote{The dronedb bars be on the left side of cloud bars}}
% \label{fig:baseline-query}
% \end{figure}

% Figure \ref{fig:baseline-query} shows how \tdblite performs against a InfluxDB cloud on spatiotemporal queries of varying dimensions. \tdblite outperforms the cloud by almost a factor of 3. We suspect that spatial indexing is to blame for the poor performance  of the cloud. On the other hand, shard-based load balancing, $MinBlock-RC$, helps improve the observed latency.`

\subsection{Mobility}
\addcb{While \tdblite is designed to operate in highly mobile environments, such as UAV fleets in disaster regions, the impact of drone mobility is confined to the communication plane. Specifically, mobility may influence the handover of drones between edge servers. However, the data plane protocols, including shard replication, spatio-temporal indexing, and distributed query execution remain unaffected by such mobility. Any mobility-induced drone unreachability may affect the freshness of query results but does not compromise the performance or correctness of the system. This separation ensures that \tdblite maintains correctness and efficiency of data operations regardless of drone movement patterns. Thus, the affect of mobility is negligible on application layer. Investigating the implications of mobility-induced communication handovers (e.g., latency or transient connection loss) is an important direction for future work, but is considered beyond the scope of this paper.}

%%%%%%%%%%%%%%%%%%%%%%%%%%%%%%%%%%%%%%%%%%%%%%%%%%%%%%%%
%%%%%%%%%%%%%%%%%%%%%%%%%%%%%%%%%%%%%%%%%%%%%%%%%%%%%%%%
\section{Conclusions}% \Note{0.25pgs}
% \ysnote{@SS: pls take a quick pass}
We have presented the design and implementation of \tdblite - a lightweight replicated database which has been optimized to process and store very large spatio-temporal data on mobile UAVs which act as edges. To the best of our knowledge, \tdblite is the first of its kind spatio-temporal datastore which can be operated on UAVs supported by a small cluster of edge nodes strategically deployed to provide emergency services in disaster regions. Using content-based hashing and indexing, coupled with a decentralized and locality-aware distributed execution engine, \tdblite is designed to provide near real-time performance in presence of network and edge failures that are atypical of disaster situations. Using containerized deployment of \tdblite over a large number of drones and edges, we demonstrate that it is able to scale well with increasing dataset size, level of concurrency, and failures. We demonstrate a 10 times improvement in performance with insertion workloads, and 100 times improvement with query workloads over the cloud baseline.

\bibliographystyle{ACM-Reference-Format}
\bibliography{references}


\begin{thebibliography}{65}

%%% the Web address.

\ifx \showCODEN    \undefined \def \showCODEN     #1{\unskip}     \fi
\ifx \showDOI      \undefined \def \showDOI       #1{#1}\fi
\ifx \showISBNx    \undefined \def \showISBNx     #1{\unskip}     \fi
\ifx \showISBNxiii \undefined \def \showISBNxiii  #1{\unskip}     \fi
\ifx \showISSN     \undefined \def \showISSN      #1{\unskip}     \fi
\ifx \showLCCN     \undefined \def \showLCCN      #1{\unskip}     \fi
\ifx \shownote     \undefined \def \shownote      #1{#1}          \fi
\ifx \showarticletitle \undefined \def \showarticletitle #1{#1}   \fi
\ifx \showURL      \undefined \def \showURL       {\relax}        \fi
% invisible to TeX
\providecommand\bibfield[2]{#2}
\providecommand\bibinfo[2]{#2}
\providecommand\natexlab[1]{#1}
\providecommand\showeprint[2][]{arXiv:#2}

\bibitem[Abraham and Roddick(1999)]%
        {abraham1999survey}
\bibfield{author}{\bibinfo{person}{Tamas Abraham} {and} \bibinfo{person}{John~F Roddick}.} \bibinfo{year}{1999}\natexlab{}.
\newblock \showarticletitle{Survey of spatio-temporal databases}.
\newblock \bibinfo{journal}{\emph{GeoInformatica}}  \bibinfo{volume}{3} (\bibinfo{year}{1999}), \bibinfo{pages}{61--99}.
\newblock


\bibitem[Allen et~al\mbox{.}(2019)]%
        {10.1145/3307334.3328627}
\bibfield{author}{\bibinfo{person}{Ryan Allen}, \bibinfo{person}{Michael Nekrasov}, {and} \bibinfo{person}{Elizabeth Belding}.} \bibinfo{year}{2019}\natexlab{}.
\newblock \showarticletitle{Data Collection from Outdoor IoT 802.15.4 Sensor Networks Using Unmanned Aerial Systems (Poster)}. In \bibinfo{booktitle}{\emph{Proceedings of the 17th Annual International Conference on Mobile Systems, Applications, and Services}} (Seoul, Republic of Korea) \emph{(\bibinfo{series}{MobiSys '19})}. \bibinfo{publisher}{Association for Computing Machinery}, \bibinfo{address}{New York, NY, USA}, \bibinfo{pages}{564–565}.
\newblock
\showISBNx{9781450366618}
\urldef\tempurl%
\url{https://doi.org/10.1145/3307334.3328627}
\showDOI{\tempurl}


\bibitem[Aral and Ovatman(2018)]%
        {aral2018decentralized}
\bibfield{author}{\bibinfo{person}{Atakan Aral} {and} \bibinfo{person}{Tolga Ovatman}.} \bibinfo{year}{2018}\natexlab{}.
\newblock \showarticletitle{A decentralized replica placement algorithm for edge computing}.
\newblock \bibinfo{journal}{\emph{IEEE transactions on network and service management}} \bibinfo{volume}{15}, \bibinfo{number}{2} (\bibinfo{year}{2018}), \bibinfo{pages}{516--529}.
\newblock


\bibitem[Asadzadeh et~al\mbox{.}(2022)]%
        {ASADZADEH2022109633}
\bibfield{author}{\bibinfo{person}{Saeid Asadzadeh}, \bibinfo{person}{Wilson~José de Oliveira}, {and} \bibinfo{person}{Carlos~Roberto de {Souza Filho}}.} \bibinfo{year}{2022}\natexlab{}.
\newblock \showarticletitle{UAV-based remote sensing for the petroleum industry and environmental monitoring: State-of-the-art and perspectives}.
\newblock \bibinfo{journal}{\emph{Journal of Petroleum Science and Engineering}}  \bibinfo{volume}{208} (\bibinfo{year}{2022}), \bibinfo{pages}{109633}.
\newblock
\showISSN{0920-4105}
\urldef\tempurl%
\url{https://doi.org/10.1016/j.petrol.2021.109633}
\showDOI{\tempurl}


\bibitem[{Bandwidth Place}(2023)]%
        {bandwidthplace2023}
\bibfield{author}{\bibinfo{person}{{Bandwidth Place}}.} \bibinfo{year}{2023}\natexlab{}.
\newblock \bibinfo{title}{Speed Comparison: 5G vs 4G LTE vs 3G}.
\newblock \bibinfo{howpublished}{\url{https://www.bandwidthplace.com/article/speed-comparison-5g-4g-lte-3g}}.
\newblock
\newblock
\shownote{Accessed: 2025-04-17}.


\bibitem[Bine et~al\mbox{.}(2022)]%
        {9773134}
\bibfield{author}{\bibinfo{person}{Lailla M.~S. Bine}, \bibinfo{person}{Azzedine Boukerche}, \bibinfo{person}{Linnyer~B. Ruiz}, {and} \bibinfo{person}{Antonio A.~F. Loureiro}.} \bibinfo{year}{2022}\natexlab{}.
\newblock \showarticletitle{Leveraging Urban Computing with the Internet of Drones}.
\newblock \bibinfo{journal}{\emph{IEEE Internet of Things Magazine}} \bibinfo{volume}{5}, \bibinfo{number}{1} (\bibinfo{year}{2022}), \bibinfo{pages}{160--165}.
\newblock
\urldef\tempurl%
\url{https://doi.org/10.1109/IOTM.003.2100091}
\showDOI{\tempurl}


\bibitem[Breitbach et~al\mbox{.}(2019)]%
        {8767386}
\bibfield{author}{\bibinfo{person}{Martin Breitbach}, \bibinfo{person}{Dominik Schäfer}, \bibinfo{person}{Janick Edinger}, {and} \bibinfo{person}{Christian Becker}.} \bibinfo{year}{2019}\natexlab{}.
\newblock \showarticletitle{Context-Aware Data and Task Placement in Edge Computing Environments}. In \bibinfo{booktitle}{\emph{2019 IEEE International Conference on Pervasive Computing and Communications (PerCom}}. \bibinfo{pages}{1--10}.
\newblock
\urldef\tempurl%
\url{https://doi.org/10.1109/PERCOM.2019.8767386}
\showDOI{\tempurl}


\bibitem[Broder and Mitzenmacher(2001)]%
        {broder2001using}
\bibfield{author}{\bibinfo{person}{Andrei Broder} {and} \bibinfo{person}{Michael Mitzenmacher}.} \bibinfo{year}{2001}\natexlab{}.
\newblock \showarticletitle{Using multiple hash functions to improve IP lookups}. In \bibinfo{booktitle}{\emph{Proceedings IEEE INFOCOM 2001. Conference on Computer Communications. Twentieth Annual Joint Conference of the IEEE Computer and Communications Society (Cat. No. 01CH37213)}}, Vol.~\bibinfo{volume}{3}. IEEE, \bibinfo{pages}{1454--1463}.
\newblock


\bibitem[Busacca et~al\mbox{.}(2020)]%
        {busacca2020drone}
\bibfield{author}{\bibinfo{person}{Fabio Busacca}, \bibinfo{person}{Laura Galluccio}, {and} \bibinfo{person}{Sergio Palazzo}.} \bibinfo{year}{2020}\natexlab{}.
\newblock \showarticletitle{Drone-assisted edge computing: a game-theoretical approach}. In \bibinfo{booktitle}{\emph{IEEE INFOCOM 2020-IEEE Conference on Computer Communications Workshops (INFOCOM WKSHPS)}}. IEEE, \bibinfo{pages}{671--676}.
\newblock


\bibitem[Chen et~al\mbox{.}(2018)]%
        {chen2018spatio}
\bibfield{author}{\bibinfo{person}{Lixing Chen}, \bibinfo{person}{Jie Xu}, \bibinfo{person}{Shaolei Ren}, {and} \bibinfo{person}{Pan Zhou}.} \bibinfo{year}{2018}\natexlab{}.
\newblock \showarticletitle{Spatio--temporal edge service placement: A bandit learning approach}.
\newblock \bibinfo{journal}{\emph{IEEE Transactions on Wireless Communications}} \bibinfo{volume}{17}, \bibinfo{number}{12} (\bibinfo{year}{2018}), \bibinfo{pages}{8388--8401}.
\newblock


\bibitem[Chen et~al\mbox{.}(2021)]%
        {9439126}
\bibfield{author}{\bibinfo{person}{Mengyu Chen}, \bibinfo{person}{Weifa Liang}, {and} \bibinfo{person}{Sajal~K. Das}.} \bibinfo{year}{2021}\natexlab{}.
\newblock \showarticletitle{Data Collection Utility Maximization in Wireless Sensor Networks via Efficient Determination of UAV Hovering Locations}. In \bibinfo{booktitle}{\emph{2021 IEEE International Conference on Pervasive Computing and Communications (PerCom)}}. \bibinfo{pages}{1--10}.
\newblock
\urldef\tempurl%
\url{https://doi.org/10.1109/PERCOM50583.2021.9439126}
\showDOI{\tempurl}


\bibitem[Christodoulou and Kolios(2020)]%
        {9128519}
\bibfield{author}{\bibinfo{person}{C. Christodoulou} {and} \bibinfo{person}{P. Kolios}.} \bibinfo{year}{2020}\natexlab{}.
\newblock \showarticletitle{Optimized tour planning for drone-based urban traffic monitoring}. In \bibinfo{booktitle}{\emph{2020 IEEE 91st Vehicular Technology Conference (VTC2020-Spring)}}. \bibinfo{pages}{1--5}.
\newblock
\urldef\tempurl%
\url{https://doi.org/10.1109/VTC2020-Spring48590.2020.9128519}
\showDOI{\tempurl}


\bibitem[Cozma et~al\mbox{.}(2022)]%
        {s22030860}
\bibfield{author}{\bibinfo{person}{Alexandru Cozma}, \bibinfo{person}{Adrian-Cosmin Firculescu}, \bibinfo{person}{Dan Tudose}, {and} \bibinfo{person}{Laura Ruse}.} \bibinfo{year}{2022}\natexlab{}.
\newblock \showarticletitle{Autonomous Multi-Rotor Aerial Platform for Air Pollution Monitoring}.
\newblock \bibinfo{journal}{\emph{Sensors}} \bibinfo{volume}{22}, \bibinfo{number}{3} (\bibinfo{year}{2022}).
\newblock
\showISSN{1424-8220}
\urldef\tempurl%
\url{https://www.mdpi.com/1424-8220/22/3/860}
\showURL{%
\tempurl}


\bibitem[De~Groot et~al\mbox{.}(2023)]%
        {10099247}
\bibfield{author}{\bibinfo{person}{Lucan De~Groot}, \bibinfo{person}{Talia Xu}, {and} \bibinfo{person}{Marco~Zúñiga Zamalloa}.} \bibinfo{year}{2023}\natexlab{}.
\newblock \showarticletitle{DroneVLC: Exploiting Drones and VLC to Gather Data from Batteryless Sensors}. In \bibinfo{booktitle}{\emph{2023 IEEE International Conference on Pervasive Computing and Communications (PerCom)}}. \bibinfo{pages}{242--251}.
\newblock
\urldef\tempurl%
\url{https://doi.org/10.1109/PERCOM56429.2023.10099247}
\showDOI{\tempurl}


\bibitem[DeCandia et~al\mbox{.}(2007)]%
        {decandia2007dynamo}
\bibfield{author}{\bibinfo{person}{Giuseppe DeCandia}, \bibinfo{person}{Deniz Hastorun}, \bibinfo{person}{Madan Jampani}, \bibinfo{person}{Gunavardhan Kakulapati}, \bibinfo{person}{Avinash Lakshman}, \bibinfo{person}{Alex Pilchin}, \bibinfo{person}{Swaminathan Sivasubramanian}, \bibinfo{person}{Peter Vosshall}, {and} \bibinfo{person}{Werner Vogels}.} \bibinfo{year}{2007}\natexlab{}.
\newblock \showarticletitle{Dynamo: Amazon's highly available key-value store}.
\newblock \bibinfo{journal}{\emph{ACM SIGOPS operating systems review}} \bibinfo{volume}{41}, \bibinfo{number}{6} (\bibinfo{year}{2007}), \bibinfo{pages}{205--220}.
\newblock


\bibitem[Erdelj et~al\mbox{.}(2017)]%
        {7807176}
\bibfield{author}{\bibinfo{person}{Milan Erdelj}, \bibinfo{person}{Enrico Natalizio}, \bibinfo{person}{Kaushik~R. Chowdhury}, {and} \bibinfo{person}{Ian~F. Akyildiz}.} \bibinfo{year}{2017}\natexlab{}.
\newblock \showarticletitle{Help from the Sky: Leveraging UAVs for Disaster Management}.
\newblock \bibinfo{journal}{\emph{IEEE Pervasive Computing}} \bibinfo{volume}{16}, \bibinfo{number}{1} (\bibinfo{year}{2017}), \bibinfo{pages}{24--32}.
\newblock
\urldef\tempurl%
\url{https://doi.org/10.1109/MPRV.2017.11}
\showDOI{\tempurl}


\bibitem[Fortune({[n.\,d.]})]%
        {doi:10.1142/9789814355858_0006}
\bibfield{author}{\bibinfo{person}{Steven Fortune}.} \bibinfo{year}{[n.\,d.]}\natexlab{}.
\newblock \bibinfo{booktitle}{\emph{Voronoi Diagram and Delaunay Triangulations}}.
\newblock \bibinfo{pages}{193--233}.
\newblock
\urldef\tempurl%
\url{https://doi.org/10.1142/9789814355858_0006}
\showDOI{\tempurl}


\bibitem[Fortune(1987)]%
        {10.1007/BF01840357}
\bibfield{author}{\bibinfo{person}{Steven Fortune}.} \bibinfo{year}{1987}\natexlab{}.
\newblock \showarticletitle{A Sweepline Algorithm for Voronoi Diagrams}.
\newblock \bibinfo{journal}{\emph{Algorithmica}} \bibinfo{volume}{2}, \bibinfo{number}{1–4} (\bibinfo{date}{nov} \bibinfo{year}{1987}), \bibinfo{pages}{153–174}.
\newblock
\showISSN{0178-4617}
\urldef\tempurl%
\url{https://doi.org/10.1007/BF01840357}
\showDOI{\tempurl}


\bibitem[Garg et~al\mbox{.}(2020a)]%
        {garg2020torquedb}
\bibfield{author}{\bibinfo{person}{Dhruv Garg}, \bibinfo{person}{Prathik Shirolkar}, \bibinfo{person}{Anshu Shukla}, {and} \bibinfo{person}{Yogesh Simmhan}.} \bibinfo{year}{2020}\natexlab{a}.
\newblock \showarticletitle{Torquedb: Distributed querying of time-series data from edge-local storage}. In \bibinfo{booktitle}{\emph{Euro-Par 2020: Parallel Processing: 26th International Conference on Parallel and Distributed Computing, Warsaw, Poland, August 24--28, 2020, Proceedings 26}}. Springer, \bibinfo{pages}{281--295}.
\newblock


\bibitem[Garg et~al\mbox{.}(2020b)]%
        {10.1007/978-3-030-57675-2_18}
\bibfield{author}{\bibinfo{person}{Dhruv Garg}, \bibinfo{person}{Prathik Shirolkar}, \bibinfo{person}{Anshu Shukla}, {and} \bibinfo{person}{Yogesh Simmhan}.} \bibinfo{year}{2020}\natexlab{b}.
\newblock \showarticletitle{TorqueDB: Distributed Querying of Time-Series Data from Edge-local Storage}. In \bibinfo{booktitle}{\emph{Euro-Par 2020: Parallel Processing}}, \bibfield{editor}{\bibinfo{person}{Maciej Malawski} {and} \bibinfo{person}{Krzysztof Rzadca}} (Eds.). \bibinfo{publisher}{Springer International Publishing}, \bibinfo{address}{Cham}, \bibinfo{pages}{281--295}.
\newblock
\showISBNx{978-3-030-57675-2}


\bibitem[Ham et~al\mbox{.}(2016)]%
        {ham2016visual}
\bibfield{author}{\bibinfo{person}{Youngjib Ham}, \bibinfo{person}{Kevin~K Han}, \bibinfo{person}{Jacob~J Lin}, {and} \bibinfo{person}{Mani Golparvar-Fard}.} \bibinfo{year}{2016}\natexlab{}.
\newblock \showarticletitle{Visual monitoring of civil infrastructure systems via camera-equipped Unmanned Aerial Vehicles (UAVs): a review of related works}.
\newblock \bibinfo{journal}{\emph{Visualization in Engineering}} \bibinfo{volume}{4}, \bibinfo{number}{1} (\bibinfo{year}{2016}), \bibinfo{pages}{1--8}.
\newblock


\bibitem[Hansen et~al\mbox{.}(2020)]%
        {hansen2020content}
\bibfield{author}{\bibinfo{person}{Casper Hansen}, \bibinfo{person}{Christian Hansen}, \bibinfo{person}{Jakob~Grue Simonsen}, \bibinfo{person}{Stephen Alstrup}, {and} \bibinfo{person}{Christina Lioma}.} \bibinfo{year}{2020}\natexlab{}.
\newblock \showarticletitle{Content-aware neural hashing for cold-start recommendation}. In \bibinfo{booktitle}{\emph{Proceedings of the 43rd International ACM SIGIR Conference on Research and Development in Information Retrieval}}. \bibinfo{pages}{971--980}.
\newblock


\bibitem[He et~al\mbox{.}(2020)]%
        {10.1145/3386901.3388912}
\bibfield{author}{\bibinfo{person}{Songtao He}, \bibinfo{person}{Favyen Bastani}, \bibinfo{person}{Arjun Balasingam}, \bibinfo{person}{Karthik Gopalakrishna}, \bibinfo{person}{Ziwen Jiang}, \bibinfo{person}{Mohammad Alizadeh}, \bibinfo{person}{Hari Balakrishnan}, \bibinfo{person}{Michael Cafarella}, \bibinfo{person}{Tim Kraska}, {and} \bibinfo{person}{Sam Madden}.} \bibinfo{year}{2020}\natexlab{}.
\newblock \showarticletitle{BeeCluster: Drone Orchestration via Predictive Optimization}. In \bibinfo{booktitle}{\emph{Proceedings of the 18th International Conference on Mobile Systems, Applications, and Services}} (Toronto, Ontario, Canada) \emph{(\bibinfo{series}{MobiSys '20})}. \bibinfo{publisher}{Association for Computing Machinery}, \bibinfo{address}{New York, NY, USA}, \bibinfo{pages}{299–311}.
\newblock
\showISBNx{9781450379540}
\urldef\tempurl%
\url{https://doi.org/10.1145/3386901.3388912}
\showDOI{\tempurl}


\bibitem[Huang et~al\mbox{.}(2013)]%
        {4glatencypaper}
\bibfield{author}{\bibinfo{person}{Junxian Huang}, \bibinfo{person}{Feng Qian}, \bibinfo{person}{Yihua Guo}, \bibinfo{person}{Yuanyuan Zhou}, \bibinfo{person}{Qiang Xu}, \bibinfo{person}{Z.~Morley Mao}, \bibinfo{person}{Subhabrata Sen}, {and} \bibinfo{person}{Oliver Spatscheck}.} \bibinfo{year}{2013}\natexlab{}.
\newblock \showarticletitle{An in-depth study of LTE: effect of network protocol and application behavior on performance}. In \bibinfo{booktitle}{\emph{Proceedings of the ACM SIGCOMM 2013 Conference on SIGCOMM}} (Hong Kong, China) \emph{(\bibinfo{series}{SIGCOMM '13})}. \bibinfo{publisher}{Association for Computing Machinery}, \bibinfo{address}{New York, NY, USA}, \bibinfo{pages}{363–374}.
\newblock
\showISBNx{9781450320566}
\urldef\tempurl%
\url{https://doi.org/10.1145/2486001.2486006}
\showDOI{\tempurl}


\bibitem[Jaiswal and Sidhanta(2022)]%
        {9665659}
\bibfield{author}{\bibinfo{person}{Shashwat Jaiswal} {and} \bibinfo{person}{Subhajit Sidhanta}.} \bibinfo{year}{2022}\natexlab{}.
\newblock \showarticletitle{Toward a smart multi-unmanned aerial vehicle system}.
\newblock \bibinfo{journal}{\emph{IEEE Potentials}} \bibinfo{volume}{41}, \bibinfo{number}{1} (\bibinfo{year}{2022}), \bibinfo{pages}{22--25}.
\newblock
\urldef\tempurl%
\url{https://doi.org/10.1109/MPOT.2021.3117259}
\showDOI{\tempurl}


\bibitem[Jeong et~al\mbox{.}(2017)]%
        {jeong2017mobile}
\bibfield{author}{\bibinfo{person}{Seongah Jeong}, \bibinfo{person}{Osvaldo Simeone}, {and} \bibinfo{person}{Joonhyuk Kang}.} \bibinfo{year}{2017}\natexlab{}.
\newblock \showarticletitle{Mobile edge computing via a UAV-mounted cloudlet: Optimization of bit allocation and path planning}.
\newblock \bibinfo{journal}{\emph{IEEE Transactions on Vehicular Technology}} \bibinfo{volume}{67}, \bibinfo{number}{3} (\bibinfo{year}{2017}), \bibinfo{pages}{2049--2063}.
\newblock


\bibitem[Jha et~al\mbox{.}(2021)]%
        {10.1145/3447993.3483273}
\bibfield{author}{\bibinfo{person}{Sagar Jha}, \bibinfo{person}{Youjie Li}, \bibinfo{person}{Shadi Noghabi}, \bibinfo{person}{Vaishnavi Ranganathan}, \bibinfo{person}{Peeyush Kumar}, \bibinfo{person}{Andrew Nelson}, \bibinfo{person}{Michael Toelle}, \bibinfo{person}{Sudipta Sinha}, \bibinfo{person}{Ranveer Chandra}, {and} \bibinfo{person}{Anirudh Badam}.} \bibinfo{year}{2021}\natexlab{}.
\newblock \showarticletitle{Visage: Enabling Timely Analytics for Drone Imagery}. In \bibinfo{booktitle}{\emph{Proceedings of the 27th Annual International Conference on Mobile Computing and Networking}} (New Orleans, Louisiana) \emph{(\bibinfo{series}{MobiCom '21})}. \bibinfo{publisher}{Association for Computing Machinery}, \bibinfo{address}{New York, NY, USA}, \bibinfo{pages}{789–803}.
\newblock
\showISBNx{9781450383424}
\urldef\tempurl%
\url{https://doi.org/10.1145/3447993.3483273}
\showDOI{\tempurl}


\bibitem[Khochare et~al\mbox{.}(2021a)]%
        {DBLP:conf/infocom/KhochareSS021}
\bibfield{author}{\bibinfo{person}{Aakash Khochare}, \bibinfo{person}{Yogesh Simmhan}, \bibinfo{person}{Francesco~Betti Sorbelli}, {and} \bibinfo{person}{Sajal~K. Das}.} \bibinfo{year}{2021}\natexlab{a}.
\newblock \showarticletitle{Heuristic Algorithms for Co-scheduling of Edge Analytics and Routes for {UAV} Fleet Missions}. In \bibinfo{booktitle}{\emph{40th {IEEE} Conference on Computer Communications, {INFOCOM} 2021, Vancouver, BC, Canada, May 10-13, 2021}}. \bibinfo{publisher}{{IEEE}}, \bibinfo{pages}{1--10}.
\newblock
\urldef\tempurl%
\url{https://doi.org/10.1109/INFOCOM42981.2021.9488740}
\showDOI{\tempurl}


\bibitem[Khochare et~al\mbox{.}(2021b)]%
        {9488740}
\bibfield{author}{\bibinfo{person}{Aakash Khochare}, \bibinfo{person}{Yogesh Simmhan}, \bibinfo{person}{Francesco~Betti Sorbelli}, {and} \bibinfo{person}{Sajal~K. Das}.} \bibinfo{year}{2021}\natexlab{b}.
\newblock \showarticletitle{Heuristic Algorithms for Co-scheduling of Edge Analytics and Routes for UAV Fleet Missions}. In \bibinfo{booktitle}{\emph{IEEE INFOCOM 2021 - IEEE Conference on Computer Communications}}. \bibinfo{pages}{1--10}.
\newblock
\urldef\tempurl%
\url{https://doi.org/10.1109/INFOCOM42981.2021.9488740}
\showDOI{\tempurl}


\bibitem[Kolomvatsos et~al\mbox{.}(2019)]%
        {kolomvatsos2019spatio}
\bibfield{author}{\bibinfo{person}{Kostas Kolomvatsos}, \bibinfo{person}{Panagiota Papadopoulou}, \bibinfo{person}{Christos Anagnostopoulos}, {and} \bibinfo{person}{Stathes Hadjiefthymiades}.} \bibinfo{year}{2019}\natexlab{}.
\newblock \showarticletitle{A Spatio-temporal data imputation model for supporting analytics at the edge}. In \bibinfo{booktitle}{\emph{Digital Transformation for a Sustainable Society in the 21st Century: 18th IFIP WG 6.11 Conference on e-Business, e-Services, and e-Society, I3E 2019, Trondheim, Norway, September 18--20, 2019, Proceedings 18}}. Springer, \bibinfo{pages}{138--150}.
\newblock


\bibitem[Koubarakis et~al\mbox{.}(2012)]%
        {koubarakis2012teleios}
\bibfield{author}{\bibinfo{person}{Manolis Koubarakis}, \bibinfo{person}{Mihai Datcu}, \bibinfo{person}{Charalambos Kontoes}, \bibinfo{person}{Ugo Di~Giammatteo}, \bibinfo{person}{Stefan Manegold}, {and} \bibinfo{person}{Eva Klien}.} \bibinfo{year}{2012}\natexlab{}.
\newblock \showarticletitle{TELEIOS: a database-powered virtual earth observatory}.
\newblock \bibinfo{journal}{\emph{Proceedings of the VLDB Endowment}} \bibinfo{volume}{5}, \bibinfo{number}{12} (\bibinfo{year}{2012}), \bibinfo{pages}{2010--2013}.
\newblock


\bibitem[Krishna et~al\mbox{.}(2023)]%
        {krishna2023using}
\bibfield{author}{\bibinfo{person}{RKN~Sai Krishna}, \bibinfo{person}{Chandrasekhar Tekur}, \bibinfo{person}{Ramesh Bhashyam}, \bibinfo{person}{Venkat Nannaka}, {and} \bibinfo{person}{Ravi Mukkamala}.} \bibinfo{year}{2023}\natexlab{}.
\newblock \showarticletitle{Using Cuckoo Filters to Improve Performance in Object Store-based Very Large Databases}. In \bibinfo{booktitle}{\emph{2023 IEEE 13th Annual Computing and Communication Workshop and Conference (CCWC)}}. IEEE, \bibinfo{pages}{0795--0800}.
\newblock


\bibitem[Kyrkou et~al\mbox{.}(2019)]%
        {8598647}
\bibfield{author}{\bibinfo{person}{Christos Kyrkou}, \bibinfo{person}{Stelios Timotheou}, \bibinfo{person}{Panayiotis Kolios}, \bibinfo{person}{Theocharis Theocharides}, {and} \bibinfo{person}{Christos Panayiotou}.} \bibinfo{year}{2019}\natexlab{}.
\newblock \showarticletitle{Drones: Augmenting Our Quality of Life}.
\newblock \bibinfo{journal}{\emph{IEEE Potentials}} \bibinfo{volume}{38}, \bibinfo{number}{1} (\bibinfo{year}{2019}), \bibinfo{pages}{30--36}.
\newblock
\urldef\tempurl%
\url{https://doi.org/10.1109/MPOT.2018.2850386}
\showDOI{\tempurl}


\bibitem[Linaje et~al\mbox{.}(2019)]%
        {linaje2019mist}
\bibfield{author}{\bibinfo{person}{Marino Linaje}, \bibinfo{person}{Javier Berrocal}, {and} \bibinfo{person}{Alfonso Galan-Benitez}.} \bibinfo{year}{2019}\natexlab{}.
\newblock \showarticletitle{Mist and edge storage: Fair storage distribution in sensor networks}.
\newblock \bibinfo{journal}{\emph{IEEE Access}}  \bibinfo{volume}{7} (\bibinfo{year}{2019}), \bibinfo{pages}{123860--123876}.
\newblock


\bibitem[Loi et~al\mbox{.}(2013)]%
        {loi2013vlsh}
\bibfield{author}{\bibinfo{person}{Tieu~Lin Loi}, \bibinfo{person}{Jae-Pil Heo}, \bibinfo{person}{Junghwan Lee}, {and} \bibinfo{person}{Sung-eui Yoon}.} \bibinfo{year}{2013}\natexlab{}.
\newblock \showarticletitle{VLSH: Voronoi-based locality sensitive hashing}. In \bibinfo{booktitle}{\emph{2013 IEEE/RSJ International Conference on Intelligent Robots and Systems}}. IEEE, \bibinfo{pages}{5345--5352}.
\newblock


\bibitem[Monga et~al\mbox{.}(2019a)]%
        {monga2019elfstore}
\bibfield{author}{\bibinfo{person}{Sumit~Kumar Monga}, \bibinfo{person}{Sheshadri~K Ramachandra}, {and} \bibinfo{person}{Yogesh Simmhan}.} \bibinfo{year}{2019}\natexlab{a}.
\newblock \showarticletitle{ElfStore: A resilient data storage service for federated edge and fog resources}. In \bibinfo{booktitle}{\emph{2019 IEEE International Conference on Web Services (ICWS)}}. IEEE, \bibinfo{pages}{336--345}.
\newblock


\bibitem[Monga et~al\mbox{.}(2019b)]%
        {8818408}
\bibfield{author}{\bibinfo{person}{Sumit~Kumar Monga}, \bibinfo{person}{Sheshadri~K. Ramachandra}, {and} \bibinfo{person}{Yogesh Simmhan}.} \bibinfo{year}{2019}\natexlab{b}.
\newblock \showarticletitle{ElfStore: A Resilient Data Storage Service for Federated Edge and Fog Resources}. In \bibinfo{booktitle}{\emph{2019 IEEE International Conference on Web Services (ICWS)}}. \bibinfo{pages}{336--345}.
\newblock
\urldef\tempurl%
\url{https://doi.org/10.1109/ICWS.2019.00062}
\showDOI{\tempurl}


\bibitem[Moninger et~al\mbox{.}(2003)]%
        {moninger2003automated}
\bibfield{author}{\bibinfo{person}{William~R Moninger}, \bibinfo{person}{Richard~D Mamrosh}, {and} \bibinfo{person}{Patricia~M Pauley}.} \bibinfo{year}{2003}\natexlab{}.
\newblock \showarticletitle{Automated meteorological reports from commercial aircraft}.
\newblock \bibinfo{journal}{\emph{Bulletin of the American Meteorological Society}} \bibinfo{volume}{84}, \bibinfo{number}{2} (\bibinfo{year}{2003}), \bibinfo{pages}{203--216}.
\newblock


\bibitem[Mortazavi et~al\mbox{.}(2020)]%
        {9355826}
\bibfield{author}{\bibinfo{person}{Seyed~Hossein Mortazavi}, \bibinfo{person}{Mohammad Salehe}, \bibinfo{person}{Moshe Gabel}, {and} \bibinfo{person}{Eyal~de Lara}.} \bibinfo{year}{2020}\natexlab{}.
\newblock \showarticletitle{Feather: Hierarchical Querying for the Edge}. In \bibinfo{booktitle}{\emph{2020 IEEE/ACM Symposium on Edge Computing (SEC)}}. \bibinfo{pages}{271--284}.
\newblock
\urldef\tempurl%
\url{https://doi.org/10.1109/SEC50012.2020.00039}
\showDOI{\tempurl}


\bibitem[Neumann et~al\mbox{.}(2011)]%
        {neumann2011stacee}
\bibfield{author}{\bibinfo{person}{Dirk Neumann}, \bibinfo{person}{Christian Bodenstein}, \bibinfo{person}{Omer~F Rana}, {and} \bibinfo{person}{Ruby Krishnaswamy}.} \bibinfo{year}{2011}\natexlab{}.
\newblock \showarticletitle{STACEE: Enhancing storage clouds using edge devices}. In \bibinfo{booktitle}{\emph{Proceedings of the 1st ACM/IEEE workshop on Autonomic computing in economics}}. \bibinfo{pages}{19--26}.
\newblock


\bibitem[Nicoara et~al\mbox{.}(2015)]%
        {nicoara2015hermes}
\bibfield{author}{\bibinfo{person}{Daniel Nicoara}, \bibinfo{person}{Shahin Kamali}, \bibinfo{person}{Khuzaima Daudjee}, {and} \bibinfo{person}{Lei Chen}.} \bibinfo{year}{2015}\natexlab{}.
\newblock \showarticletitle{Hermes: Dynamic Partitioning for Distributed Social Network Graph Databases.}. In \bibinfo{booktitle}{\emph{EDBT}}. \bibinfo{pages}{25--36}.
\newblock


\bibitem[Nicolaescu et~al\mbox{.}(2021)]%
        {nicolaescu2021store}
\bibfield{author}{\bibinfo{person}{Adrian-Cristian Nicolaescu}, \bibinfo{person}{Spyridon Mastorakis}, {and} \bibinfo{person}{Ioannis Psaras}.} \bibinfo{year}{2021}\natexlab{}.
\newblock \showarticletitle{Store edge networked data (SEND): A data and performance driven edge storage framework}. In \bibinfo{booktitle}{\emph{IEEE INFOCOM 2021-IEEE Conference on Computer Communications}}. IEEE, \bibinfo{pages}{1--10}.
\newblock


\bibitem[Noor et~al\mbox{.}(2018)]%
        {noor2018remote}
\bibfield{author}{\bibinfo{person}{Norzailawati~Mohd Noor}, \bibinfo{person}{Alias Abdullah}, {and} \bibinfo{person}{Mazlan Hashim}.} \bibinfo{year}{2018}\natexlab{}.
\newblock \showarticletitle{Remote sensing UAV/drones and its applications for urban areas: A review}. In \bibinfo{booktitle}{\emph{IOP conference series: Earth and environmental science}}, Vol.~\bibinfo{volume}{169}. IOP Publishing, \bibinfo{pages}{012003}.
\newblock


\bibitem[Oikawa and Kondo(2021)]%
        {9439127}
\bibfield{author}{\bibinfo{person}{Hiroki Oikawa} {and} \bibinfo{person}{Masaaki Kondo}.} \bibinfo{year}{2021}\natexlab{}.
\newblock \showarticletitle{Density-Based Data Selection and Management for Edge Computing}. In \bibinfo{booktitle}{\emph{2021 IEEE International Conference on Pervasive Computing and Communications (PerCom)}}. \bibinfo{pages}{1--11}.
\newblock
\urldef\tempurl%
\url{https://doi.org/10.1109/PERCOM50583.2021.9439127}
\showDOI{\tempurl}


\bibitem[Pant et~al\mbox{.}(2018)]%
        {pant2018survey}
\bibfield{author}{\bibinfo{person}{Neelabh Pant}, \bibinfo{person}{Mohammadhani Fouladgar}, \bibinfo{person}{Ramez Elmasri}, {and} \bibinfo{person}{Kulsawasd Jitkajornwanich}.} \bibinfo{year}{2018}\natexlab{}.
\newblock \showarticletitle{A survey of spatio-temporal database research}. In \bibinfo{booktitle}{\emph{Intelligent Information and Database Systems: 10th Asian Conference, ACIIDS 2018, Dong Hoi City, Vietnam, March 19-21, 2018, Proceedings, Part II 10}}. Springer, \bibinfo{pages}{115--126}.
\newblock


\bibitem[PELEKIS et~al\mbox{.}(2004)]%
        {pelekis_theodoulidis_kopanakis_theodoridis_2004}
\bibfield{author}{\bibinfo{person}{NIKOS PELEKIS}, \bibinfo{person}{BABIS THEODOULIDIS}, \bibinfo{person}{IOANNIS KOPANAKIS}, {and} \bibinfo{person}{YANNIS THEODORIDIS}.} \bibinfo{year}{2004}\natexlab{}.
\newblock \showarticletitle{Literature review of spatio-temporal database models}.
\newblock \bibinfo{journal}{\emph{The Knowledge Engineering Review}} \bibinfo{volume}{19}, \bibinfo{number}{3} (\bibinfo{year}{2004}), \bibinfo{pages}{235–274}.
\newblock
\urldef\tempurl%
\url{https://doi.org/10.1017/S026988890400013X}
\showDOI{\tempurl}


\bibitem[Qiao et~al\mbox{.}(2020)]%
        {qiao2020trustworthy}
\bibfield{author}{\bibinfo{person}{Fuli Qiao}, \bibinfo{person}{Jun Wu}, \bibinfo{person}{Jianhua Li}, \bibinfo{person}{Ali~Kashif Bashir}, \bibinfo{person}{Shahid Mumtaz}, {and} \bibinfo{person}{Usman Tariq}.} \bibinfo{year}{2020}\natexlab{}.
\newblock \showarticletitle{Trustworthy edge storage orchestration in intelligent transportation systems using reinforcement learning}.
\newblock \bibinfo{journal}{\emph{IEEE Transactions on Intelligent Transportation Systems}} \bibinfo{volume}{22}, \bibinfo{number}{7} (\bibinfo{year}{2020}), \bibinfo{pages}{4443--4456}.
\newblock


\bibitem[Raj et~al\mbox{.}(2023)]%
        {10171496}
\bibfield{author}{\bibinfo{person}{Suman Raj}, \bibinfo{person}{Harshil Gupta}, {and} \bibinfo{person}{Yogesh Simmhan}.} \bibinfo{year}{2023}\natexlab{}.
\newblock \showarticletitle{Scheduling DNN Inferencing on Edge and Cloud for Personalized UAV Fleets}. In \bibinfo{booktitle}{\emph{2023 IEEE/ACM 23rd International Symposium on Cluster, Cloud and Internet Computing (CCGrid)}}. \bibinfo{pages}{615--626}.
\newblock
\urldef\tempurl%
\url{https://doi.org/10.1109/CCGrid57682.2023.00063}
\showDOI{\tempurl}


\bibitem[Rohi et~al\mbox{.}(2020)]%
        {ROHI2020e03252}
\bibfield{author}{\bibinfo{person}{Godall Rohi}, \bibinfo{person}{O'tega Ejofodomi}, {and} \bibinfo{person}{Godswill Ofualagba}.} \bibinfo{year}{2020}\natexlab{}.
\newblock \showarticletitle{Autonomous monitoring, analysis, and countering of air pollution using environmental drones}.
\newblock \bibinfo{journal}{\emph{Heliyon}} \bibinfo{volume}{6}, \bibinfo{number}{1} (\bibinfo{year}{2020}), \bibinfo{pages}{e03252}.
\newblock
\showISSN{2405-8440}
\urldef\tempurl%
\url{https://doi.org/10.1016/j.heliyon.2020.e03252}
\showDOI{\tempurl}


\bibitem[Schalk and Herrmann(2017)]%
        {speed_claims}
\bibfield{author}{\bibinfo{person}{Lukas~Marcel Schalk} {and} \bibinfo{person}{Martin Herrmann}.} \bibinfo{year}{2017}\natexlab{}.
\newblock \showarticletitle{Suitability of LTE for drone-to-infrastructure communications in very low level airspace}. In \bibinfo{booktitle}{\emph{2017 IEEE/AIAA 36th Digital Avionics Systems Conference (DASC)}}. \bibinfo{pages}{1--7}.
\newblock
\urldef\tempurl%
\url{https://doi.org/10.1109/DASC.2017.8102112}
\showDOI{\tempurl}


\bibitem[Sonbol et~al\mbox{.}(2020)]%
        {sonbol2020edgekv}
\bibfield{author}{\bibinfo{person}{Karim Sonbol}, \bibinfo{person}{{\"O}znur {\"O}zkasap}, \bibinfo{person}{Ibrahim Al-Oqily}, {and} \bibinfo{person}{Moayad Aloqaily}.} \bibinfo{year}{2020}\natexlab{}.
\newblock \showarticletitle{EdgeKV: decentralized, scalable, and consistent storage for the edge}.
\newblock \bibinfo{journal}{\emph{J. Parallel and Distrib. Comput.}}  \bibinfo{volume}{144} (\bibinfo{year}{2020}), \bibinfo{pages}{28--40}.
\newblock


\bibitem[Sziroczak et~al\mbox{.}(2022)]%
        {SZIROCZAK2022100859}
\bibfield{author}{\bibinfo{person}{David Sziroczak}, \bibinfo{person}{Daniel Rohacs}, {and} \bibinfo{person}{Jozsef Rohacs}.} \bibinfo{year}{2022}\natexlab{}.
\newblock \showarticletitle{Review of using small UAV based meteorological measurements for road weather management}.
\newblock \bibinfo{journal}{\emph{Progress in Aerospace Sciences}}  \bibinfo{volume}{134} (\bibinfo{year}{2022}), \bibinfo{pages}{100859}.
\newblock
\showISSN{0376-0421}
\urldef\tempurl%
\url{https://doi.org/10.1016/j.paerosci.2022.100859}
\showDOI{\tempurl}


\bibitem[Tao and Papadias(2002)]%
        {tao2002time}
\bibfield{author}{\bibinfo{person}{Yufei Tao} {and} \bibinfo{person}{Dimitris Papadias}.} \bibinfo{year}{2002}\natexlab{}.
\newblock \showarticletitle{Time-parameterized queries in spatio-temporal databases}. In \bibinfo{booktitle}{\emph{Proceedings of the 2002 ACM SIGMOD international conference on Management of data}}. \bibinfo{pages}{334--345}.
\newblock


\bibitem[Technologies(2019)]%
        {akamai}
\bibfield{author}{\bibinfo{person}{Akamai Technologies}.} \bibinfo{year}{2019}\natexlab{}.
\newblock \bibinfo{title}{\href{https://www.akamai.com/products/serverless-computing-edgeworkers}{EdgeWorkers: Create functions at the edge on the largest distributed serverless network}}.
\newblock
\newblock


\bibitem[Vasisht et~al\mbox{.}(2017)]%
        {farmbeats}
\bibfield{author}{\bibinfo{person}{Deepak Vasisht}, \bibinfo{person}{Zerina Kapetanovic}, \bibinfo{person}{Jongho Won}, \bibinfo{person}{Xinxin Jin}, \bibinfo{person}{Ranveer Chandra}, \bibinfo{person}{Sudipta Sinha}, \bibinfo{person}{Ashish Kapoor}, \bibinfo{person}{Madhusudhan Sudarshan}, {and} \bibinfo{person}{Sean Stratman}.} \bibinfo{year}{2017}\natexlab{}.
\newblock \showarticletitle{{FarmBeats}: An {IoT} Platform for {Data-Driven} Agriculture}. In \bibinfo{booktitle}{\emph{14th USENIX Symposium on Networked Systems Design and Implementation (NSDI 17)}}. \bibinfo{publisher}{USENIX Association}, \bibinfo{address}{Boston, MA}, \bibinfo{pages}{515--529}.
\newblock
\showISBNx{978-1-931971-37-9}
\urldef\tempurl%
\url{https://www.usenix.org/conference/nsdi17/technical-sessions/presentation/vasisht}
\showURL{%
\tempurl}


\bibitem[Wang et~al\mbox{.}(2023)]%
        {10.1145/3589775}
\bibfield{author}{\bibinfo{person}{Chen Wang}, \bibinfo{person}{Jialin Qiao}, \bibinfo{person}{Xiangdong Huang}, \bibinfo{person}{Shaoxu Song}, \bibinfo{person}{Haonan Hou}, \bibinfo{person}{Tian Jiang}, \bibinfo{person}{Lei Rui}, \bibinfo{person}{Jianmin Wang}, {and} \bibinfo{person}{Jiaguang Sun}.} \bibinfo{year}{2023}\natexlab{}.
\newblock \showarticletitle{Apache IoTDB: A Time Series Database for IoT Applications}.
\newblock \bibinfo{journal}{\emph{Proc. ACM Manag. Data}} \bibinfo{volume}{1}, \bibinfo{number}{2}, Article \bibinfo{articleno}{195} (\bibinfo{date}{jun} \bibinfo{year}{2023}), \bibinfo{numpages}{27}~pages.
\newblock


\bibitem[Weil et~al\mbox{.}(2004)]%
        {weil2004dynamic}
\bibfield{author}{\bibinfo{person}{Sage~A Weil}, \bibinfo{person}{Kristal~T Pollack}, \bibinfo{person}{Scott~A Brandt}, {and} \bibinfo{person}{Ethan~L Miller}.} \bibinfo{year}{2004}\natexlab{}.
\newblock \showarticletitle{Dynamic metadata management for petabyte-scale file systems}. In \bibinfo{booktitle}{\emph{SC'04: Proceedings of the 2004 ACM/IEEE conference on Supercomputing}}. IEEE, \bibinfo{pages}{4--4}.
\newblock


\bibitem[Xia et~al\mbox{.}(2019)]%
        {xia2019secure}
\bibfield{author}{\bibinfo{person}{Junxu Xia}, \bibinfo{person}{Geyao Cheng}, \bibinfo{person}{Siyuan Gu}, {and} \bibinfo{person}{Deke Guo}.} \bibinfo{year}{2019}\natexlab{}.
\newblock \showarticletitle{Secure and trust-oriented edge storage for internet of things}.
\newblock \bibinfo{journal}{\emph{IEEE Internet of Things Journal}} \bibinfo{volume}{7}, \bibinfo{number}{5} (\bibinfo{year}{2019}), \bibinfo{pages}{4049--4060}.
\newblock


\bibitem[Xu et~al\mbox{.}(2020)]%
        {9166731}
\bibfield{author}{\bibinfo{person}{Jianwen Xu}, \bibinfo{person}{Kaoru Ota}, {and} \bibinfo{person}{Mianxiong Dong}.} \bibinfo{year}{2020}\natexlab{}.
\newblock \showarticletitle{Big Data on the Fly: UAV-Mounted Mobile Edge Computing for Disaster Management}.
\newblock \bibinfo{journal}{\emph{IEEE Transactions on Network Science and Engineering}} \bibinfo{volume}{7}, \bibinfo{number}{4} (\bibinfo{year}{2020}), \bibinfo{pages}{2620--2630}.
\newblock
\urldef\tempurl%
\url{https://doi.org/10.1109/TNSE.2020.3016569}
\showDOI{\tempurl}


\bibitem[Yu et~al\mbox{.}(2020)]%
        {8956055}
\bibfield{author}{\bibinfo{person}{Zhe Yu}, \bibinfo{person}{Yanmin Gong}, \bibinfo{person}{Shimin Gong}, {and} \bibinfo{person}{Yuanxiong Guo}.} \bibinfo{year}{2020}\natexlab{}.
\newblock \showarticletitle{Joint Task Offloading and Resource Allocation in UAV-Enabled Mobile Edge Computing}.
\newblock \bibinfo{journal}{\emph{IEEE Internet of Things Journal}} \bibinfo{volume}{7}, \bibinfo{number}{4} (\bibinfo{year}{2020}), \bibinfo{pages}{3147--3159}.
\newblock
\urldef\tempurl%
\url{https://doi.org/10.1109/JIOT.2020.2965898}
\showDOI{\tempurl}


\bibitem[Yuan et~al\mbox{.}(2021)]%
        {yuan2021csedge}
\bibfield{author}{\bibinfo{person}{Liang Yuan}, \bibinfo{person}{Qiang He}, \bibinfo{person}{Feifei Chen}, \bibinfo{person}{Jun Zhang}, \bibinfo{person}{Lianyong Qi}, \bibinfo{person}{Xiaolong Xu}, \bibinfo{person}{Yang Xiang}, {and} \bibinfo{person}{Yun Yang}.} \bibinfo{year}{2021}\natexlab{}.
\newblock \showarticletitle{Csedge: Enabling collaborative edge storage for multi-access edge computing based on blockchain}.
\newblock \bibinfo{journal}{\emph{IEEE Transactions on Parallel and Distributed Systems}} \bibinfo{volume}{33}, \bibinfo{number}{8} (\bibinfo{year}{2021}), \bibinfo{pages}{1873--1887}.
\newblock


\bibitem[Zalipynis(2018)]%
        {zalipynis2018chronosdb}
\bibfield{author}{\bibinfo{person}{Ramon Antonio~Rodriges Zalipynis}.} \bibinfo{year}{2018}\natexlab{}.
\newblock \showarticletitle{Chronosdb: distributed, file based, geospatial array dbms}.
\newblock \bibinfo{journal}{\emph{Proceedings of the VLDB Endowment}} \bibinfo{volume}{11}, \bibinfo{number}{10} (\bibinfo{year}{2018}), \bibinfo{pages}{1247--1261}.
\newblock


\bibitem[Zhang et~al\mbox{.}(2021)]%
        {10.1145/3447993.3448628}
\bibfield{author}{\bibinfo{person}{Wuyang Zhang}, \bibinfo{person}{Zhezhi He}, \bibinfo{person}{Luyang Liu}, \bibinfo{person}{Zhenhua Jia}, \bibinfo{person}{Yunxin Liu}, \bibinfo{person}{Marco Gruteser}, \bibinfo{person}{Dipankar Raychaudhuri}, {and} \bibinfo{person}{Yanyong Zhang}.} \bibinfo{year}{2021}\natexlab{}.
\newblock \showarticletitle{Elf: Accelerate High-Resolution Mobile Deep Vision with Content-Aware Parallel Offloading}. In \bibinfo{booktitle}{\emph{Proceedings of the 27th Annual International Conference on Mobile Computing and Networking}} (New Orleans, Louisiana) \emph{(\bibinfo{series}{MobiCom '21})}. \bibinfo{publisher}{Association for Computing Machinery}, \bibinfo{address}{New York, NY, USA}, \bibinfo{pages}{201–214}.
\newblock
\showISBNx{9781450383424}
\urldef\tempurl%
\url{https://doi.org/10.1145/3447993.3448628}
\showDOI{\tempurl}


\bibitem[Zhao et~al\mbox{.}(2016)]%
        {zhao2016toward}
\bibfield{author}{\bibinfo{person}{Dongfang Zhao}, \bibinfo{person}{Kan Qiao}, \bibinfo{person}{Zhou Zhou}, \bibinfo{person}{Tonglin Li}, \bibinfo{person}{Zhihan Lu}, {and} \bibinfo{person}{Xiaohua Xu}.} \bibinfo{year}{2016}\natexlab{}.
\newblock \showarticletitle{Toward efficient and flexible metadata indexing of big data systems}.
\newblock \bibinfo{journal}{\emph{IEEE Transactions on Big Data}} \bibinfo{volume}{3}, \bibinfo{number}{1} (\bibinfo{year}{2016}), \bibinfo{pages}{107--117}.
\newblock


\bibitem[Zim\'{a}nyi et~al\mbox{.}(2020)]%
        {10.1145/3406534}
\bibfield{author}{\bibinfo{person}{Esteban Zim\'{a}nyi}, \bibinfo{person}{Mahmoud Sakr}, {and} \bibinfo{person}{Arthur Lesuisse}.} \bibinfo{year}{2020}\natexlab{}.
\newblock \showarticletitle{MobilityDB: A Mobility Database Based on PostgreSQL and PostGIS}.
\newblock \bibinfo{journal}{\emph{ACM Trans. Database Syst.}} \bibinfo{volume}{45}, \bibinfo{number}{4}, Article \bibinfo{articleno}{19} (\bibinfo{date}{dec} \bibinfo{year}{2020}), \bibinfo{numpages}{42}~pages.
\newblock
\showISSN{0362-5915}


\end{thebibliography}

% %% For numbered reference style

%   \textit{\LaTeX: a document preparation system},
%   Addison Wesley, Massachusetts,

% \end{thebibliography}
\end{document}